\documentclass[journal]{IEEEtran}
\usepackage[switch]{lineno}
\usepackage{framed}
\usepackage{listings}
\usepackage{xcolor}
\usepackage{graphicx}
\usepackage{caption}
\usepackage{amsmath}
\usepackage{amsfonts} 
\usepackage{algorithm}
\usepackage{algpseudocode}
\usepackage{bbding}
\usepackage{pifont}
\usepackage{amssymb}
\usepackage[top=2cm, bottom=2cm, left=2cm, right=2cm]{geometry}
\usepackage{hyperref}
\usepackage{environ}
\usepackage{tikz}
\usepackage{ragged2e}
\usepackage{hyperref}
\usepackage{booktabs}
\usepackage{multicol}
\usepackage{multirow}
\usepackage{threeparttable}
\usepackage{comment}
\usepackage{colortbl}
\usepackage{textcomp}
\usepackage{bbding}
\usepackage{CJKutf8}
\makeatletter

\newcommand{\Rmnum}[1]{\expandafter\@slowromancap\romannumeral #1@}
\makeatother
\definecolor{c1}{HTML}{7e0f12}
\NewEnviron{elaboration}{
\par
\begin{tikzpicture}
\node[rectangle,minimum width=0.48\textwidth] (m) {\begin{minipage}{0.46\textwidth}\BODY\end{minipage}};
\draw[dashed] (m.south west) rectangle (m.north east);
\end{tikzpicture}
}
\hypersetup{
    colorlinks=true,
    linkcolor=blue, 
    citecolor=blue,   
    urlcolor=black,    
}

\ifCLASSOPTIONcompsoc
\else
\fi

\ifCLASSINFOpdf
\else
\fi
\let\oldequation\equation
\let\oldendequation\endequation

\renewenvironment{equation}
  {\linenomathNonumbers\oldequation}
  {\oldendequation\endlinenomath}
\begin{document}
%
% paper title
% can use linebreaks \\ within to get better formatting as desired
% Do not put math or special symbols in the title.
\title{Don't Confuse! Redrawing GUI Navigation Flow in Mobile Apps for Visually Impaired Users}

\author{Mengxi~Zhang\dag \ddag,
        Huaxiao~Liu\dag \ddag *,
        Yuheng~Zhou\dag,
        Chunyang~Chen$\dotplus$,
        Pei~Huang\S,
		 and Jian~Zhao$\divideontimes$% <-this % stops a space
\IEEEcompsocitemizethanks{\IEEEcompsocthanksitem \dag Mengxi Zhang, Huaxiao Liu (*corresponding author) and Yuheng Zhou are with the College of Computer Science and Technology, Jilin University, Changchun, 130012, China.\protect\\
% note need leading \protect in front of \\ to get a newline within \thanks as
% \\ is fragile and will error, could use \hfil\break instead.
E-mail: zmx19@mails.jlu.edu.cn, liuhuaxiao@jlu.edu.cn, zhouyh5521@mails.jlu.edu.cn
\IEEEcompsocthanksitem \ddag Mengxi Zhang and Huaxiao Liu are also with the Key Laboratory of Symbolic Computation and Knowledge Engineering of Ministry of Education, Jilin University, Changchun, 130012, China.
\IEEEcompsocthanksitem $\dotplus$Chunyang Chen is a full professor in the Department of Computer Science, Technical University of Munich, Heilbronn, Weipertstr.8-10, 74076, Germany. E-mail: chun-yang.chen@tum.de
\IEEEcompsocthanksitem \S Pei~Huang is with the Department of Computer Science, Stanford University, California, 9025, U.S.A. E-mail: huangpei@stanford.edu
\IEEEcompsocthanksitem $\divideontimes$Jian~Zhao is with the Department of Computer Science, Changchun University, Changchun, 1340012, China, E-mail: zhaojian@ccu.edu.cn}
%\IEEEcompsocthanksitem Yuzhou Liu is with the College of Computer Science and Technology, Jilin University, Changchun, China.}
\thanks{}}

% The paper headers
\markboth{IEEE Transactions on Software Engineering}%
{Zhang \MakeLowercase{\textit{et al.}}: Don't Confuse! Redrawing GUI Navigation Flow in Mobile Apps for Visually Impaired Users}
% The only time the second header will appear is for the odd numbered pages
% after the title page when using the twoside option.
% 
% *** Note that you probably will NOT want to include the author's ***
% *** name in the headers of peer review papers.                   ***
% You can use \ifCLASSOPTIONpeerreview for conditional compilation here if
% you desire.

% use for special paper notices
%\IEEEspecialpapernotice{(Invited Paper)}

% for Computer Society papers, we must declare the abstract and index terms
% PRIOR to the title within the \IEEEtitleabstractindextext IEEEtran
% command as these need to go into the title area created by \maketitle.
% As a general rule, do not put math, special symbols or citations
% in the abstract or keywords.
% Note that keywords are not normally used for peerreview papers.

% make the title area

\maketitle

\begin{abstract}
Mobile applications (apps) are integral to our daily lives, offering diverse services and functionalities.
They enable sighted users to access information coherently in an extremely convenient manner.
However, it remains unclear if visually impaired users, who rely solely on the screen readers (e.g., Talkback) to navigate and access app information, can do so in the correct and reasonable order.
This may result in significant information bias and operational errors.
Furthermore, in our preliminary exploration, we explained and clarified that the navigation sequence-related issues encountered by visually impaired users could be categorized into two types: unintuitive navigation sequence and unapparent focus switching.
Considering these issues, in this work, we proposed a method named RGNF (\underline{\textbf{R}}e-draw \underline{\textbf{G}}UI \underline{\textbf{N}}avigation \underline{\textbf{F}}low).
It aimed to enhance the understandability and coherence of accessing the content of each component within the Graphical User Interface (GUI), together with assisting developers in creating well-designed GUI navigation flow (GNF).
This method was inspired by the characteristics identified in our preliminary study, where visually impaired users expected navigation to be associated with close position and similar shape of GUI components that were read consecutively. 
Thus, our method relied on the principles derived from the Gestalt psychological model, aiming to group GUI components into different regions according to the laws of proximity and similarity, thereby redrawing the GNFs.
To evaluate the effectiveness of our method, we calculated sequence similarity values before and after redrawing the GNF, and further employed the tools proposed by Alotaibi et al. to measure the reachability of GUI components. 
Our results demonstrated a substantial improvement in similarity (0.921) compared to the baseline (0.624), together with the reachability (90.31\%) compared to the baseline GNF (74.35\%). 
Furthermore, a qualitative user study revealed that our method had a positive effect on providing visually impaired users with an improved user experience.
\end{abstract}

%%
%% The code below is generated by the tool at http://dl.acm.org/ccs.cfm.
%% Please copy and paste the code instead of the example below.
%%

%%
%% Keywords. The author(s) should pick words that accurately describe
%% the work being presented. Separate the keywords with commas.
\begin{IEEEkeywords}
GUI, accessibility, Gestalt psychological model, navigation
\end{IEEEkeywords}

\section{Introduction}\label{sec: introduction}

\IEEEPARstart{I}{n} the increasingly digitized world, mobile applications (apps) have played an indispensable role and changed the way we communicate, work and access information~\cite{Mack2021WhatDW, Vendome2019CanEU}. 
However, as people take advantage of the convenience and efficiency of these apps, it is necessary to think about the inclusivity of this technology~\cite{10338828, Zhang2023EnhancingAO}. 
Regarding such inclusivity, the accessibility of apps is a pressing issue that requires innovative solutions to ensure that all users, regardless of their physical abilities, can fully engage with these technologies. 

Visually impaired users typically operate GUIs using screen readers (e.g., Talkback and Voice-over) in the following manner. 
First, they activate the screen reader with the system in accessibility mode, where focusable areas appear within the GUI to help locate and access GUI components. Then, users usually employ three types of interaction gestures to operate the screen reader. 
These include a single-finger touch gesture to focus on and have the screen reader read out the component being touched, a single-finger swipe gesture left or right to move the focus to the previous or next element and have it read, and a single-finger double-tap gesture to activate the currently focused component. 
With these interactive actions, visually impaired users can effectively access and navigate the various components and information within the GUIs.

\begin{figure}
\centering
\includegraphics[width=6.7cm]{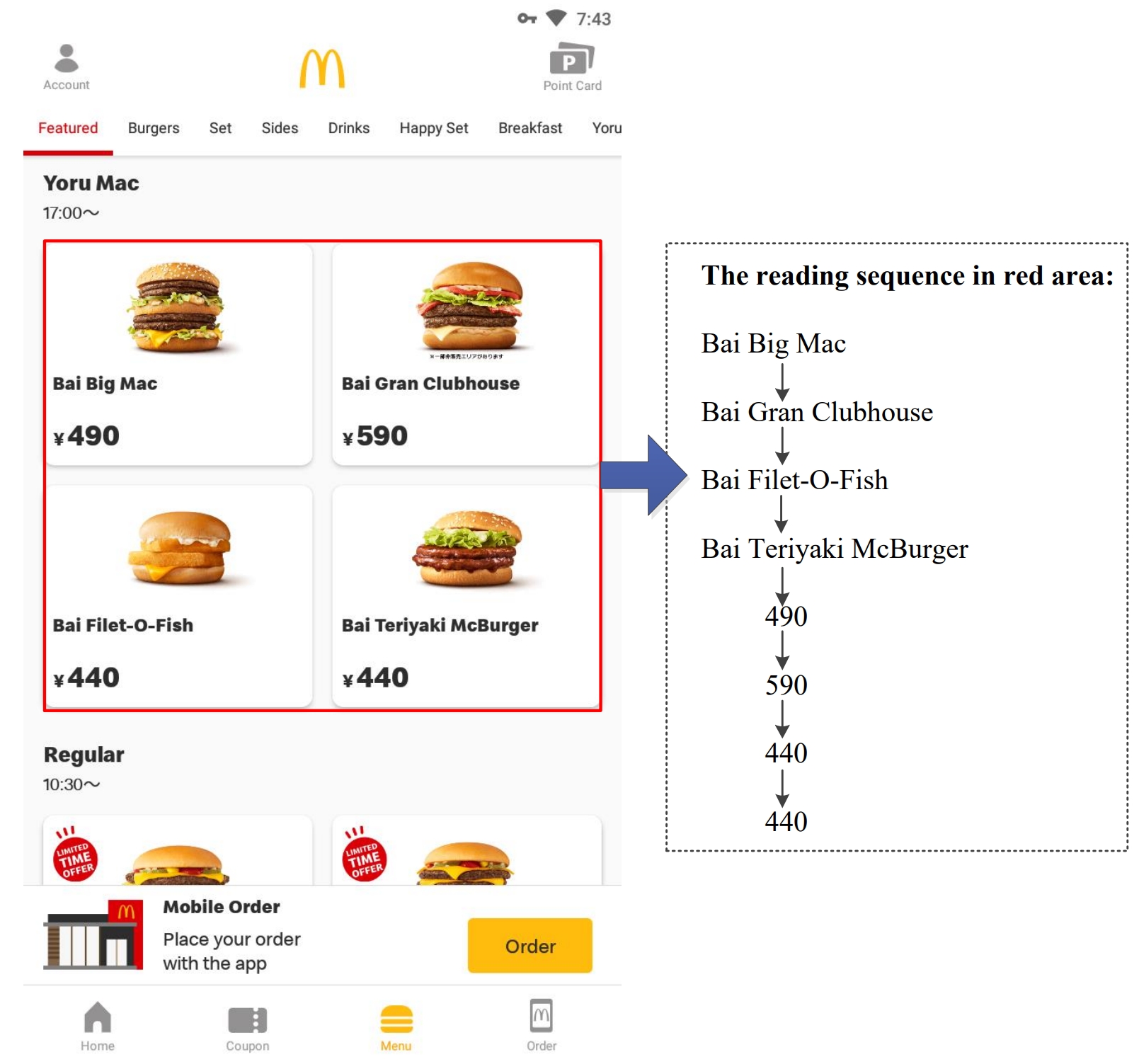}
\caption{An example of problematic navigation flow.}
\label{fig: chotic}
\end{figure}
A crucial aspect of accessibility is the Graphical User Interface (GUI) navigation flow (GNF), which refers to ``the GUI elements or components that need to be arranged in a sensible order to make it easier for users to navigate, find content and find out where it is located'' (Web Content Accessibility Guideline (WCAG) 2.1, Guideline 2.4 Navigable)~\footnote{\url{https://www.w3.org/TR/WCAG21/\#navigable}}.
If this flow is poorly structured or inefficient, users may have difficulty in understanding the logical process of GUIs, leading to confusion and reduced usability. 
Figure~\ref{fig: chotic} shows an example of problematic navigation flow. 
In this instance, the navigation of the red area is ``Bai Big Mac $\rightarrow$ Bai Gran Clubhouse $\rightarrow$ Bai Filet-O-Fish $\rightarrow$ Bai Teriyaki McBurger $\rightarrow$ 490 $\rightarrow$ 590 $\rightarrow$ 440 $\rightarrow$ 440''. 
That is to say, screen readers would first read the names of all products, and then read their prices.
This situation could confuse visually impaired users who depend exclusively on screen readers for accessing information. 
It may hinder their ability to discern which product corresponds to which price, and thus affect their completion of subsequent operations.

Considering such problems, reorganizing and optimizing the GNF would have a profound impact on app accessibility. 
Also, as we discussed in the investigation in Section~\ref{sub: Pre}, there is an obvious difference between the GNFs perceived by visually impaired users and the default results.
However, there is a relative lack of research on GNFs. 
By 2022, Alotaibi et al. firstly proposed approaches to detect issues within the GNFs and provide suggestions for developers, including the reachability of target GUI components~\cite{Chiou2023BAGELAA} and the occurrence of abrupt changes in the navigation flow~\cite{Alotaibi2022AutomatedDO}. 
Although these methods can detect the issues that related to GNFs, they fall short of providing practical solutions, leaving the resolution of these issues to the developers. 
Furthermore, there is another challenge, how should the GNF be adjusted to better serve visually impaired users, that requires consideration.
The key to overcoming this challenge is to find out how the GNF is perceived by visually impaired users, and thus strike a delicate balance between the adjusted GNFs and their desired outcomes.
This requires investigating the practical perceptions and interaction preferences of visually impaired users, in order to guide the redrawing of GNFs.

Leveraging this insight, in this paper, we proposed a method for redrawing the GUI navigation flow, named RGNF (\underline{\textbf{R}}edraw \underline{\textbf{G}}UI \underline{\textbf{N}}avigation \underline{\textbf{F}}low). 
In the beginning, through our preliminary research in Section~\ref{sub: Pre}, we delved into the cognitive behavior of visually impaired users to understand the nature of GNFs from their perspective.
We discovered that these users tend to perceive GUI components that are close in position and similar in shape as being read consecutively.
Building upon this observation, we employed the laws of proximity and similarity from the Gestalt psychological model~\cite{Ashcraft1997FundamentalsOC, Xie2022PsychologicallyinspiredUI} to group the GUI components into regions.
Subsequently, we reorganized the GNFs of these components based on the newly defined regions.
In this regard, our method provided developers with clear guidelines for GUI navigation design, which ensured that the GNF can satisfy the expectations of visually impaired users well.

To assess our proposed approach, we conducted a series of experiments to answer two research questions. 
The first research question (RQ1) centered on evaluating the effectiveness of our method. 
To this end, we conducted comparative experiments between our redrawn GNFs and the default GNFs in terms of the similarity and reachability to user-annotated GNFs. 
In addition, we carrieed out an ablation study on the filtering step of our method. 
The outcomes of these experiments demonstrate that our redrawn GNFs have higher similarity (0.921) compared to the default GNFs (0.624), and our approach achieves an average improvement of 15.96\% (90.31\%-74.35\%) in reachability.
Moreover, the results of the ablation study show that the inclusion of the filtering step is essential. 
The second research question (RQ2) focused on assessing the practical usefulness of our approach. 
To answer this question, we conducted a user study with 8 visually impaired users. 
The findings from this user study indicated that our redrawn GNFs are well-received by users.

In summary, this work makes the following three contributions:
\begin{itemize}
\item To the best of our knowledge, this is the first work capable of adjusting and redrawing the GNFs.
\item We present a method for redesigning GNFs, which provides developers with better design guidance aligned with the operational habits of visually impaired users.
\item We open source the code and data related to this work~\cite{DataCode}, aiming to promote subsequent related research and provide data support.
\end{itemize}

The remainder of this paper is organized as follows.
Section~\ref{sec: background} explains what GUI navigation flow is and conducts preliminary research.
Section~\ref{sec: approach} introduces the main process of our proposed method.
Section~\ref{sec: experiment} evaluates the effectiveness and usefulness of our method.
Section~\ref{sec: threats} discusses the possible risks and solutions involved in the whole work.
Section~\ref{sec: related work} presents the works that are related to ours.
Section~\ref{sec: conclusion} draws the conclusion. 

\begin{figure}
\centering
\includegraphics[width=8.7cm]{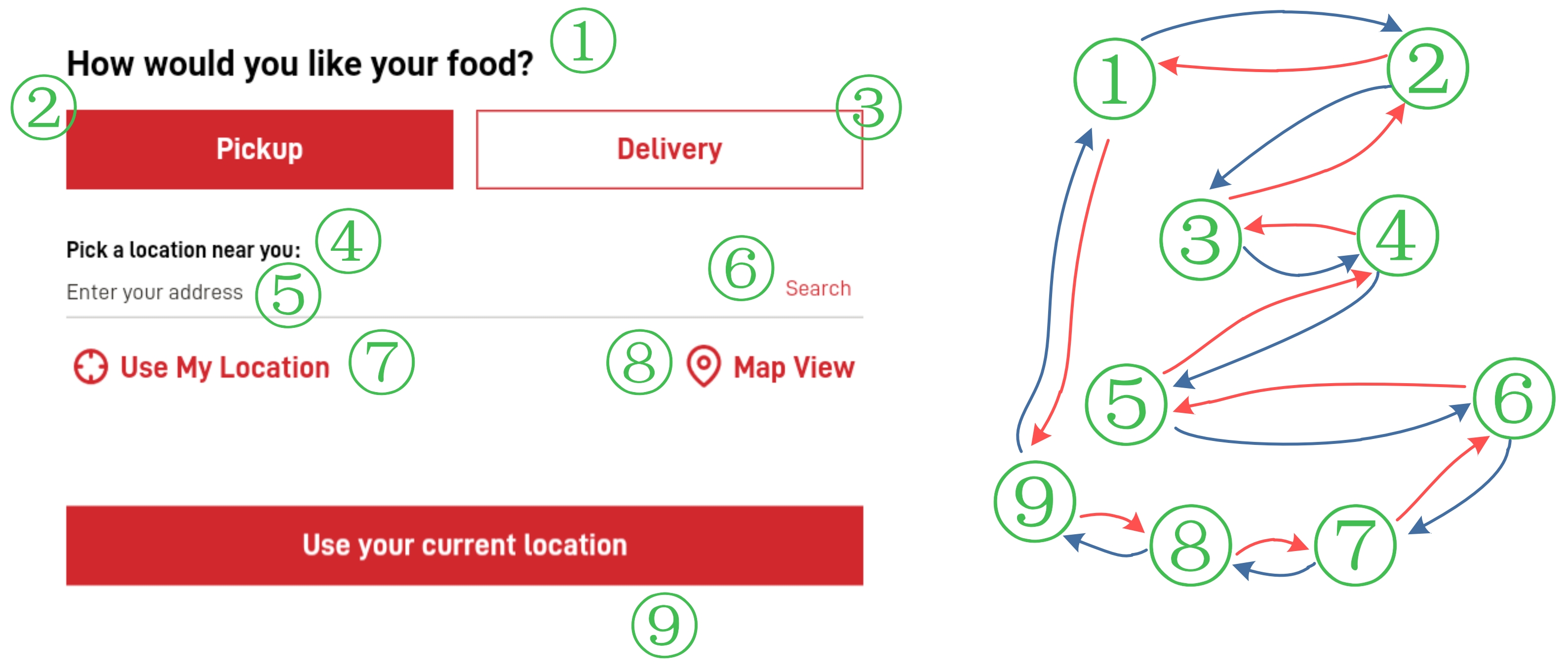}
\caption{An example of GUI navigation flow.}
\label{fig: GNF}
\end{figure}
\section{Background and Preliminary}\label{sec: background}

In this section, we first explained what a GUI navigation flow (GNF) is using a concrete example.
Then, we conducted preliminary research to find out how the GNF was usually perceived by visually impaired users, and observed the prevalent characteristics.
\subsection{GUI Navigation Flow (GNF)}

The GUI Navigation Flow (GNF) plays a crucial role in defining how visually impaired users can move through an app, access its functions, and ultimately achieve their goals~\cite{Chiou2023DetectingDK}. 
For a practical example, see Figure~\ref{fig: GNF}, Talkback reads the components within this GUI sequentially following the directions indicated by the arrows in this figure. 
In this example, the blue arrows represent the outcome of the user swiping to the right, while the red arrows indicate swiping to the left. 
When visually impaired users navigate the GUI in a sliding manner, they could swipe right to continuously access information or swipe left to return.
The sequence of components visited in turn by this sliding forms the GNF.
Notably, in the context of our study, we only address the GNFs associated with swiping to the right, without considering leftward swiping. 
Swiping left is typically associated with users intending to navigate back to specific components.
Also, the bidirectional correspondence with rightward swiping implies that if issues are identified in the GNF for rightward swiping, similar issues would naturally extend to leftward swiping as well.

\begin{figure*}
\centering
\includegraphics[width=17.7cm]{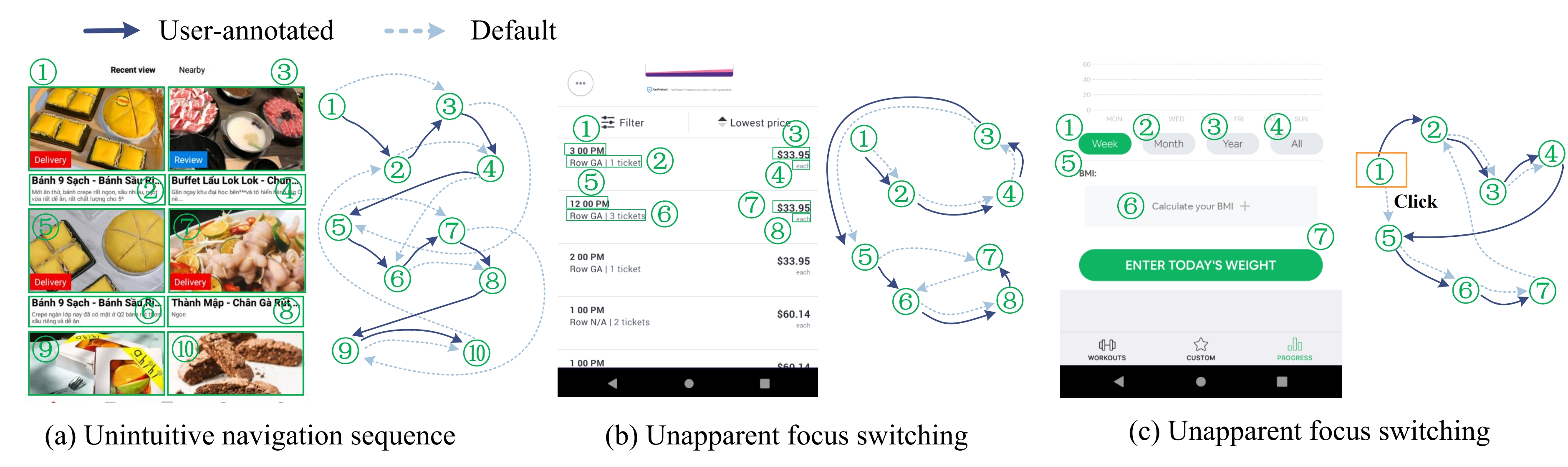}
\caption{Examples of annotated GNFs. Only swipe-right operations are presented in this context.}
\label{fig: examples}
\end{figure*}
\subsection{Preliminary Research}\label{sub: Pre}

In this research, we aim to elucidate the GNFs as perceived by visually impaired users in their cognitive processes.
Therefore, we proceed with the following three steps from the perspective of actual visually impaired users: participants recruitment, research setting, and the analysis of results.

\subsubsection{Participants recruitment}

We recruited participants from a school of special education for our research, and it was conducted through the dissemination of a recruitment notice.
In this notice, we briefly introduced the purpose of our research, the time commitment required (3-4 hours), the rewards (30\$ shopping card), the privacy protection protocol, together with having a foundation in English.
After two weeks of waiting, we received 15 user applications.
Following the exclusion of users with less than 3 years of mobile phone usage experience, we ultimately invited 10 users to participate in our study.
These participants consisted of 5 males and 5 females, with ages ranging from 18 to 27 years. 
Also, they had diverse professional backgrounds, encompassing massage therapy, music, psychology, software development, and law. 
Furthermore, we ensured that these 10 participants strictly adhered to the principles of voluntariness, and any research involving them was contingent upon obtaining their ethical consent. 
We also obtained ethical approval including informed consent, anonymity, confidentiality, voluntary participation, and respect for autonomy, to ensure the protection of participants' rights and privacy throughout the research process.
Participants were given the autonomy to withdraw from the study at any point during their involvement.

\subsubsection{Research setting}
In the initial phase of our research, we gathered apps for participants to evaluate.
This process adheres to two principles: first, to ensure that the apps cover a wide range of domains that visually impaired users might use, and second, to ensure that the number of selected apps does not impose an excessive burden on participants.
Therefore, we turned our attention to the recommended lists on Google Play, which encompassed 18 commonly used app domains, totaling 289 apps on Oct 26, 2023.
Among them, we excluded game apps from this dataset due to their tendency to feature an excessive amount of dynamic, visually intensive content, which might not be directly relevant to blind users.
As such, we assembled a dataset consisting of 265 apps from 17 domains.
From these apps, we further selected the Top-2 rated apps from each domain for the investigation.

Thus far, we have collected a total of 34 apps for this research.
Then, we invited all participants into a quiet room where each participant was accompanied by a staff member to provide the necessary assistance and recorded the results.
During the investigation, we repeated the purpose and the tasks participants were required to complete.
Next, we distributed the apps to the participants.
To mitigate the possible bias introduced by individual factors, we assigned the same app to different participants, and ensured that each app was annotated by at least two participants. 
Participants then evaluated the apps according to the following procedure: They first gained the approximate locations of components within the GUIs through touch exploration, and then utilized TalkBack to navigate the components via swipe browsing.
If the information obtained about the components did not match their expectations, they notified nearby staff members.
Throughout this process, staff members observed participants' actions and manually recorded the sequence they want to navigate, forming the user-annotated navigation flow.

After finishing the above recordings, the first three authors of this paper then participated in the process of visualizing and analyzing these results.
Initially, we visualized the GNFs in the apps involved by visually impaired users and compared them with the default navigation of the Talkback, annotating inconsistent areas.
Subsequently, we further observed the GNFs annotated by visually impaired users, including the relative positional relationships and visual differences between two consecutively read components.
During this process, the three authors simultaneously observed the same GNF and its corresponding GUI, and they then engaged in a focus group discussion~\cite{Kitzinger1995QualitativeRI} to determine the characteristics.
Notably, this analysis was conducted using observational methods, as our research results did not involve data information.

\subsubsection{Results and analysis}

Figure~\ref{fig: examples} illustrates the examples of the GNFs annotated by visually impaired users with solid blue arrows, and the default navigation of Talkback that is displayed in light blue arrows.
Among them, we observed disparities between the desired navigation process of visually impaired users and the actual navigation provided by Talkback, which constituted accessibility issues within the GNFs.
As identified in previous studies~\cite{Chiou2023BAGELAA, Alotaibi2022AutomatedDO}, the problem depicted in Figure~\ref{fig: examples} (a) refers to \emph{unintuitive navigation sequence}, indicating that the navigation sequence provided by the screen reader fails to intuitively and effectively present coherent content for visually impaired users.
This example is specifically manifested as follows.
From the users' perspective, the components \ding{172} and \ding{173} should be read consecutively, however, the Talkback default reads \ding{172} followed by \ding{174}, which is unintuitive to understand the contents.
Another category of the issue is \emph{unapparent focus switching}, akin to the sudden changes in navigation sequence direction observed in Figure~\ref{fig: examples} (b) and (c).
Seeing Figure~\ref{fig: examples} (b) for a detailed explanation, of when Talkback reads the content within the view group containing components \ding{172}, \ding{173}, \ding{174}, and \ding{175}, the sequence is ``\ding{172} $\rightarrow$ \ding{173} $\rightarrow$ \ding{174} $\rightarrow$ \ding{175}''.
However, when the focus shifts to the view group containing components \ding{176}, \ding{177}, \ding{178}, and \ding{179}, the sequence becomes ``\ding{176} $\rightarrow$ \ding{178} $\rightarrow$ \ding{177} $\rightarrow$ \ding{179}''.
As for the example in Figure~\ref{fig: examples} (c), when component \ding{172} is clicked, the focus of Talkback is suddenly switched to component \ding{176} instead of reading component \ding{173}.

Further, we explored the characteristics of the GNFs annotated by users and find two results.
The first one is related to the location of GUI components.
We observed that components positioned closer within the same container are more likely to be read consecutively, as illustrated by components \ding{176} and \ding{177}, as well as components \ding{178} and \ding{179} in Figure~\ref{fig: examples} (b).
The second characteristic comes from the shape of GUI components.
According to the annotated results, the visually similar components should also be read consecutively, such as components \ding{172}, \ding{173}, \ding{174}, and \ding{175} in Figure~\ref{fig: examples} (c).
Such observations, to some extent, indicate that visually impaired users' expected navigation also exhibits certain regularities, motivating us to seek an appropriate method to reflect their cognitive behaviors.
We also acknowledged that this small-scale survey may not fully represent all visually impaired users. 
Still, we made efforts to involve visually impaired users from diverse occupations and ages, aiming to provide insightful ideas to make the redesigned GNFs more aligned with the operational habits of these users.

Following the aforementioned characteristics, it is imperative for us to pay attention to the positions and shapes of components when redrawing the GNFs. 
To achieve this, we focused on the psychological models that reflect user behaviors, aiming to find solutions therein.

Gestalt psychological model~\cite{Ashcraft1997FundamentalsOC} is the way that we had found to represent user behavior.
It suggests that human perception and comprehension of phenomena often adhere to the principle of perceiving the whole before its parts, where the cognitive understanding of the parts depends on the nature of the whole~\cite{Ashcraft1997FundamentalsOC, Xie2022PsychologicallyinspiredUI}. 
In our work, we focused on its two perceptual laws as outlined below to direct our approach to redrawing GNFs.
\romannumeral1) \emph{Law of proximity: ``The perception of objects within a perceptual field is determined by the degree of proximity or contiguity among their respective parts. 
The closer the individual parts are to each other, the higher the likelihood of their amalgamation in perception''.}
From this law, we can infer that when users obtain information within a GUI, they typically assume that information between components positioned adjacent or nearby is related. Consequently, they tend to inspect these components together.
\romannumeral2) \emph{Law of similarity: ``When perceiving, people tend to amalgamate items with similar stimulus elements, as long as they are not disrupted by proximity factors. In other words, similar components tend to form distinct groups in perception.''}
From the perspective of user-GUI interaction, this law essentially elucidates that components with similar shapes, appearances, or types are often grouped by users. 
Consequently, users tend to gather information from these components sequentially.

\begin{figure}
\centering
\includegraphics[width=8.2cm]{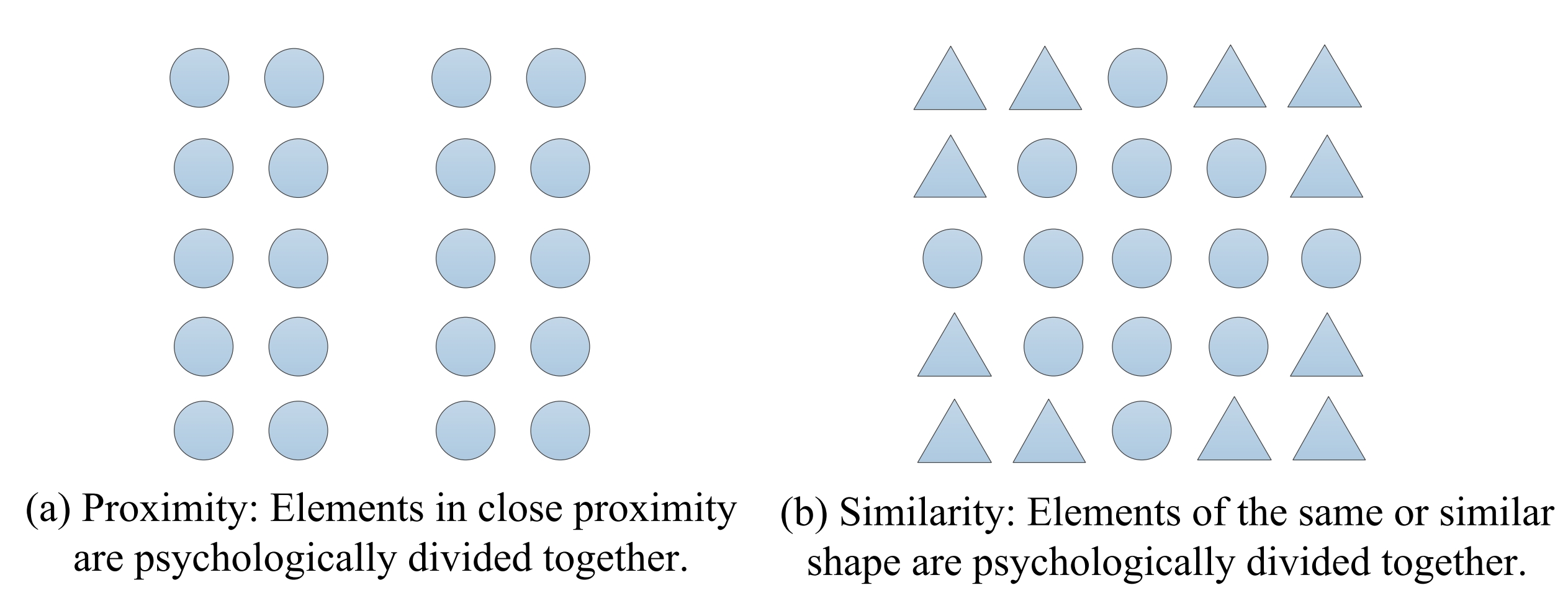}
\caption{A visual example of proximity and similarity laws in Gestalt principles.}
\label{fig: gestalt}
\end{figure}

To provide a clear understanding of the laws of proximity and similarity in Gestalt principles and their relevance to visually impaired users, we presented an example shown in Figure~\ref{fig: gestalt}. 
In Figure~\ref{fig: gestalt} (a), the visual elements were naturally divided into two regions due to psychological factors, illustrating the principle of proximity. 
This principle suggested that elements close to one another tended to be perceived as part of the same group. 
In Figure~\ref{fig: gestalt} (b), despite the elements having different shapes, we observed a tendency to group the same or similar shapes (circle and triangle) together, demonstrating the principle of similarity. 
For visually impaired users, Gestalt principles aligned with their neurovisual processing of GUI images in non-visual contexts.
Proximity, for example, helped users associate related auditory cues, while similarity provided consistency that aided in understanding functionality across components.
Furthermore, by using Gestalt principles, we could obtain insights into how GUIs were perceived and processed by the human brain. 
These principles supported our approach to grouping elements within the GUI, which subsequently allowed for the reordering of their navigation sequence, providing visually impaired users with a more user-friendly navigation process.

The scenario mentioned above essentially mirrors the principles adhered to GUI visual design~\footnote{\url{https://www.interaction-design.org/literature/topics/visual-design}}, which tended to convey coherent information through components that were positional close or visually similar. 
Consequently, our approach revolved around this viewpoint, aiming to redraw the GUI navigation process to enable visually impaired users, who cannot perceive visual cues, to seamlessly access GUI information.

\begin{figure*}
\centering
\includegraphics[width=18cm]{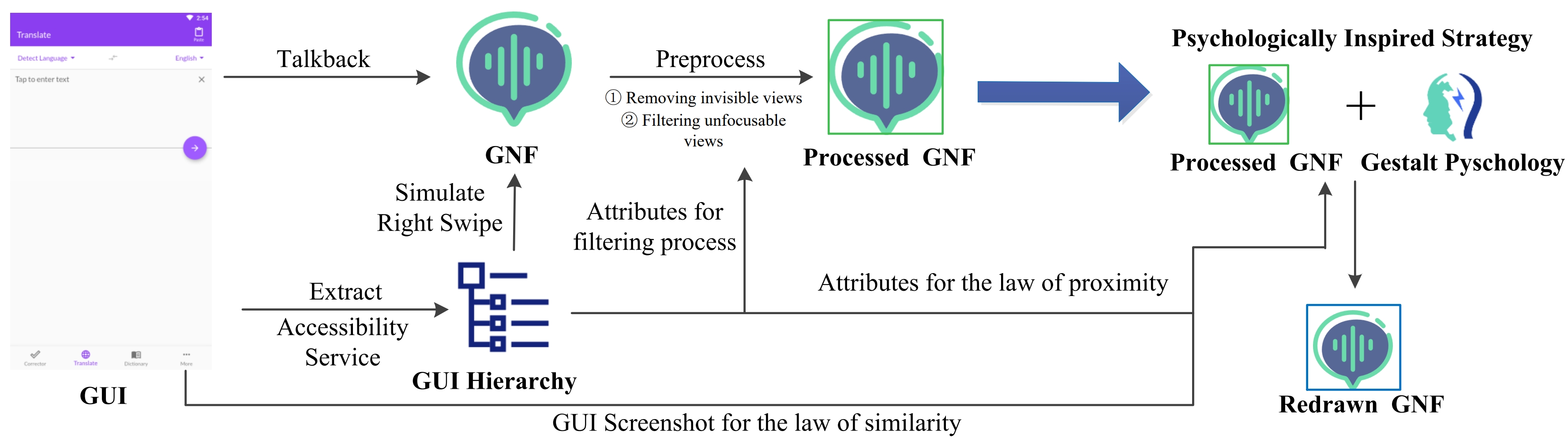}
\caption{The workflow of our approach.}
\label{fig: overview}
\end{figure*}
\section{Approach}\label{sec: approach}

In this section, we elaborated on the method proposed for redrawing the GUI navigation flow (GNF) in the following process. 
Figure~\ref{fig: overview} illustrates the primary workflow of our method. 
First, given a GUI to be processed, our method employs the Accessibility Service~\cite{AccessibilityService} to extract its hierarchy, which is represented in a tree structure.
Following that, we adopt a script to simulate the user's right swipe action within this tree, record the traversal results of Talkback and form the original GNF.
Still, due to the presence of invisible and unfocusable views in the GUI hierarchy, which might not be accessed by visually impaired users, our method further removes them and obtains the processed GNF.
In the subsequent phase, the processed GNF undergoes a redraw process inspired by the laws of proximity and similarity of the Gestalt psychological model~\cite{Ashcraft1997FundamentalsOC}. 
Such a GNF would finally be provided to developers to guide them in addressing issues therein or to aid in subsequent design endeavors.
It can also provide more coherent navigation for visually impaired users and help them to access information better.

\subsection{Extraction and Preprocessing of GNF}\label{sub: preprocess}

The first step in this section is to extract the GUI hierarchies.
To achieve this, we utilized the \emph{onAccessibilityEvent()} method provided by the Accessibility Service for extraction.
The obtained information is structured in the form of a view tree, encompassing the attributes of resource\_id, view types, texts, bounds, focusability, and enablement for each GUI component~\cite{Zhang2023AutomatedAC}. 
Subsequently, based on this view tree, we created a script capable of simulating user navigation by swiping right on the GUI.
This script iteratively visits all components within the GUI, executing right swipe operations, thereby connecting each component visited to the succeeding one and constructing the GNF.
Notably, such GNFs do not concern specific textual contents; rather, they solely record the sequential order in which GUI components are read.

The second step involves preprocessing the obtained GNF, which includes the exclusion of invisible and unfocusable views.
Invisible views refer to graphical elements or components that are present within the interface but are not currently perceptible or rendered within the user's visual field during normal interaction~\cite{Sun2023PropertyBasedFF}.
These invisible views typically exist in the form of out-of-viewport elements or empty views left by the developer to tweak the layout~\cite{Alshayban2020AccessibilityII}. 
They may encompass various interface elements, such as blank buttons, menu items, or content panels, that are positioned outside the immediate viewable area, either due to screen size limitations or the layout design of the GUI.
To remove these views, we defined a set of keyphrases that identify such views, typically associated with their corresponding attributes. 
For example, given one of the components in the layout file of a GUI to be analyzed, we focused on three types of information in this file.
This includes the ``visibility'' attribute of components, their ``class name'', and whether they can respond to user touch events or gestures.
In our recognition process, if the attribute of  ``visibility'' is marked as ``gone'' or ``invisible'', the words ``hidden-element'' or ``invisible'' appear in the class name, or the component is unable to respond to events or gestures, then we consider this component is invisible.
Notably, for visually impaired users, we considered a component to be invisible if it cannot respond to events or gestures, as it lacked any functionality in such cases, making it neither clickable nor operable.

As for the unfocusable views, which are the elements that cannot be accessed by screen readers~\cite{Alotaibi2023ScaleFixAA}.
Thus, these views cannot be navigated by users through swiping and are excluded from the constructed GNFs.
Yet, there exists a special case for such views, wherein if the ``enable'' attribute is set to ``true'' but the ``focusable'' attribute is ``false'', then the component is reachable by screen readers but unable to output its information.
Such components not only impede visually impaired users from accessing information but also disrupt the coherence of their information access. 
Hence, we filtered out the view if its attributes of ``focusable'' and ``enable'' conform to the above situation.

The deletion of the aforementioned views may consequently result in the disconnection of components in the GNFs.
To maintain the integrity of them, we performed the following supplementation. 
If the deleted component has no subsequent components, it is directly removed; however, if it serves as an intermediary connection between two other nodes is deleted, we need to connect those two nodes.
For instance, there are three GUI components $i$,$j$, and $k$ connected in turn, and if $j$ is considered unfocusable and subsequently removed, the connection between components $i$ and $k$ would be broken. 
To address this, we directly connected $i$ and $k$ to maintain the connectivity of the GNF.
In summary, the processed GNFs exhibit two main advantages: firstly, they can comprehensively reflect the navigation process of GUI components; secondly, they avoid the insertion of irrelevant information caused by invisible and unfocusable components when redrawing the navigation.

\vspace{-0.42em}
\subsection{Psychologically Inspired Strategy}\label{sub: psy}

In this section, we elaborate on how to utilize the Gestalt model to group the GUI components.
The detailed process of utilizing the laws of proximity and similarity from this psychological model is shown below:
\begin{figure}
\centering
\includegraphics[width=6.8cm]{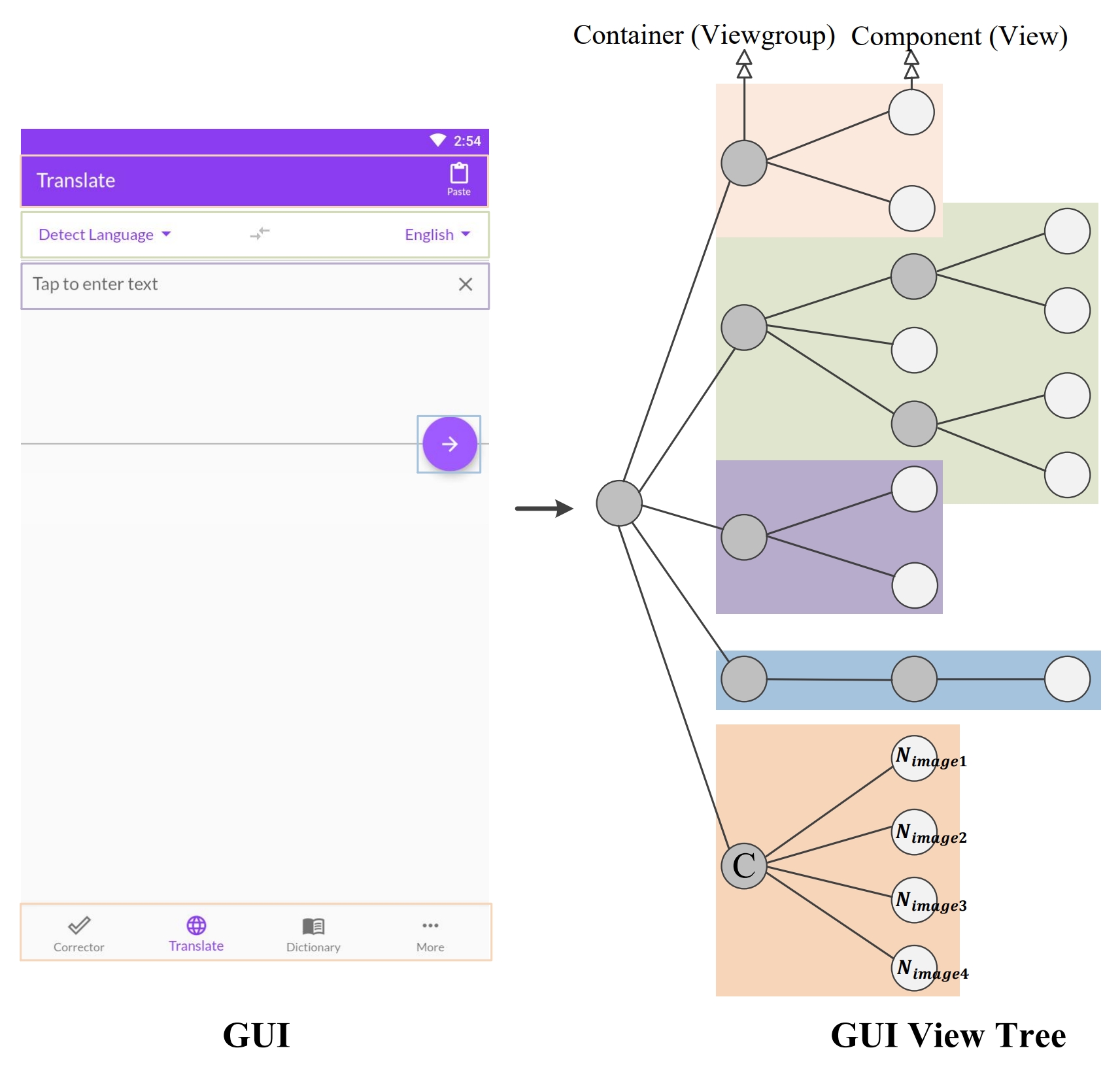}
\caption{An example of \emph{initial-regions}.}
\label{fig: proximity}
\end{figure}

\subsubsection{Proximity strategy} This law elucidates that users tend to perceive GUI components positioned nearby together.
To accommodate this tendency, we grouped the regions to which components belong based on the container they are in and their positions.
Algorithm~\ref{alg:gui_regions} shows the specific process and related pseudo-code of using proximity strategy to group regions.
At first, within the GUI hierarchy, containers typically exist in the form of \emph{$<$viewgroup$>$}, while components appear in various views, such as \emph{$<$textview$>$} and \emph{$<$imageview$>$}.
This was shown in lines 6 to 9 of algorithm~\ref{alg:gui_regions}, we started with the highest level of container nodes, grouped all the child nodes contained by each container node, and called them \emph{initial-regions}.
Specifically, as illustrated in Figure~\ref{fig: proximity}, where the \emph{initial-regions} for this example are delineated by different colors.
To provide a more detailed explanation, we took the components within the bottom navigation area (highlighted in orange of Figure~\ref{fig: proximity}) as an example, which encompasses five nodes. 
Among them, the type of node $C$ is identified as a \emph{$<$viewgroup$>$}, while $N_{image1}$, $N_{image2}$, $N_{image3}$, and $N_{image4}$ are of type \emph{$<$imageview$>$}.
As such, our approach initially grouped the four components ($N_{image1}$, $N_{image2}$, $N_{image3}$, and $N_{image4}$) into the same region.
Applying the same procedure to the other components within the GUI, we obtained the \emph{initial-regions} of the GUI components.

\begin{algorithm}
\footnotesize
\caption{Grouping Proximal Regions}
\label{alg:gui_regions}

\begin{algorithmic}[1]

\Procedure{Group\_Proximal\_Regions}{viewTree}
    \State \textbf{Input:} GUI view tree $viewTree$
    \State \textbf{Output:} Set of proximal regions $proximalRegions$
    
    \State Initialize $initialRegions \gets \emptyset$
    \State Initialize $proximalRegions \gets \emptyset$
    
    \For{each container $C_i \in viewTree$ where $C_i$ is $<$viewgroup$>$}
        \State Group all child nodes of $C_i$ into $initialRegion$
        \State Add $initialRegion$ to $initialRegions$
    \EndFor
    
    \For{each $initialRegion \in initialRegions$}
        \State Initialize $currentProximalRegion \gets \emptyset$
        
        \For{each component $c_j \in initialRegion$}
            \State Obtain $bounds_j \gets c_j.bounds$ 
            
            \For{each component $c_k \in initialRegion$ where $c_k \neq c_j$}
                \State Obtain $bounds_k \gets c_k.bounds$
                
                \State Calculate $distance \gets \textsc{CalculateDistance}(bounds_j, bounds_k)$
                
                \If{$0 \leq distance \leq 15$} 
                    \State Add $c_j$ and $c_k$ to $currentRegion$
                \EndIf
            \EndFor
            
        \EndFor
        
        \State Add $currentRegion$ to $proximalRegions$
    \EndFor
    
    \State \textbf{return} $proximalRegions$
\EndProcedure

\Function{CalculateDistance}{$j, k$}
    \State \textbf{Input:} Bounds of j[$x_j$,$y_j$][$m_j$,$n_j$], k[$x_k$,$y_k$][$m_k$,$n_k$]
    \State \textbf{Output:} Distance $distance$ between components
    
    \State $distance \gets |k.m_k-j.m_j-j.y_j|$
    
    \State \textbf{return} $distance$
\EndFunction

\end{algorithmic}
\end{algorithm}

Further, we proceeded to group the components in the same \emph{initial-region} according to their positions, and called these regions \emph{proximal-regions}.
Lines 10 to 24 of algorithm~\ref{alg:gui_regions} show this process.
The positional information of the components is obtained through the ``bounds'' attribute of each component in the view tree. 
This attribute provides the left-top corner coordinates of the GUI component's bounding, along with the length and width of the component.
Then, we established minimum and maximum thresholds to determine the proximity of the two GUI components as follows.
Following the Android developer guidelines for layouts~\footnote{\url{https://developer.android.com/guide/topics/ui/declaring-layout}}, it specifies the minimum spacing between two components is 5 pixels.
However, GUI components may not adhere strictly to this standard, as smaller spacings may be used to accommodate more content. 
Therefore, we set the minimum threshold to 0 pixels to encompass all components as far as possible. 
For the maximum threshold, we tested with the multiples of 5 pixels specified in the guidelines through grid search on the experimental datasets in Section~\ref{sub: data}. 
We found that when the spacing exceeds 15 pixels (3 times the specified value in the guideline), there might be other components between two components, resulting in a loss of proximity relationship. 
Hence, we set the maximum threshold to 15 pixels.

Still using the previous example illustrated in Figure~\ref{fig: proximity}, but shift the attention to the various components within the green region, as shown in Figure~\ref{fig: position}.
This region contains five components, named {$N_{text5}$, $N_{button6}$, $N_{button7}$, $N_{text8}$, and $N_{button9}$, and the positions for them are as follows:
($N_{text5}$:[28,256][25,545]), ($N_{button6}$:[28,28][286,545]), ($N_{button7}$:[38,38][542,545]), ($N_{text8}$:[28,128][872,545]), ($N_{button9}$:[28,28][1004,545]).
Utilizing the aforementioned thresholds, we grouped these five nodes; specifically, grouping $N_{text5}$ and $N_{button6}$ due to satisfying the condition (286-25-256=5 pixels), as well as $N_{text8}$ and $N_{button9}$.
The $N_{button7}$ is separately assigned to a region because its position with other components all exceeds the maximum threshold. 
The final \emph{proximal-regions} for the GUI components are illustrated in Figure~\ref{fig: position}, together with components of different results are delineated by distinct colors.
\begin{figure}
\centering
\includegraphics[width=5cm]{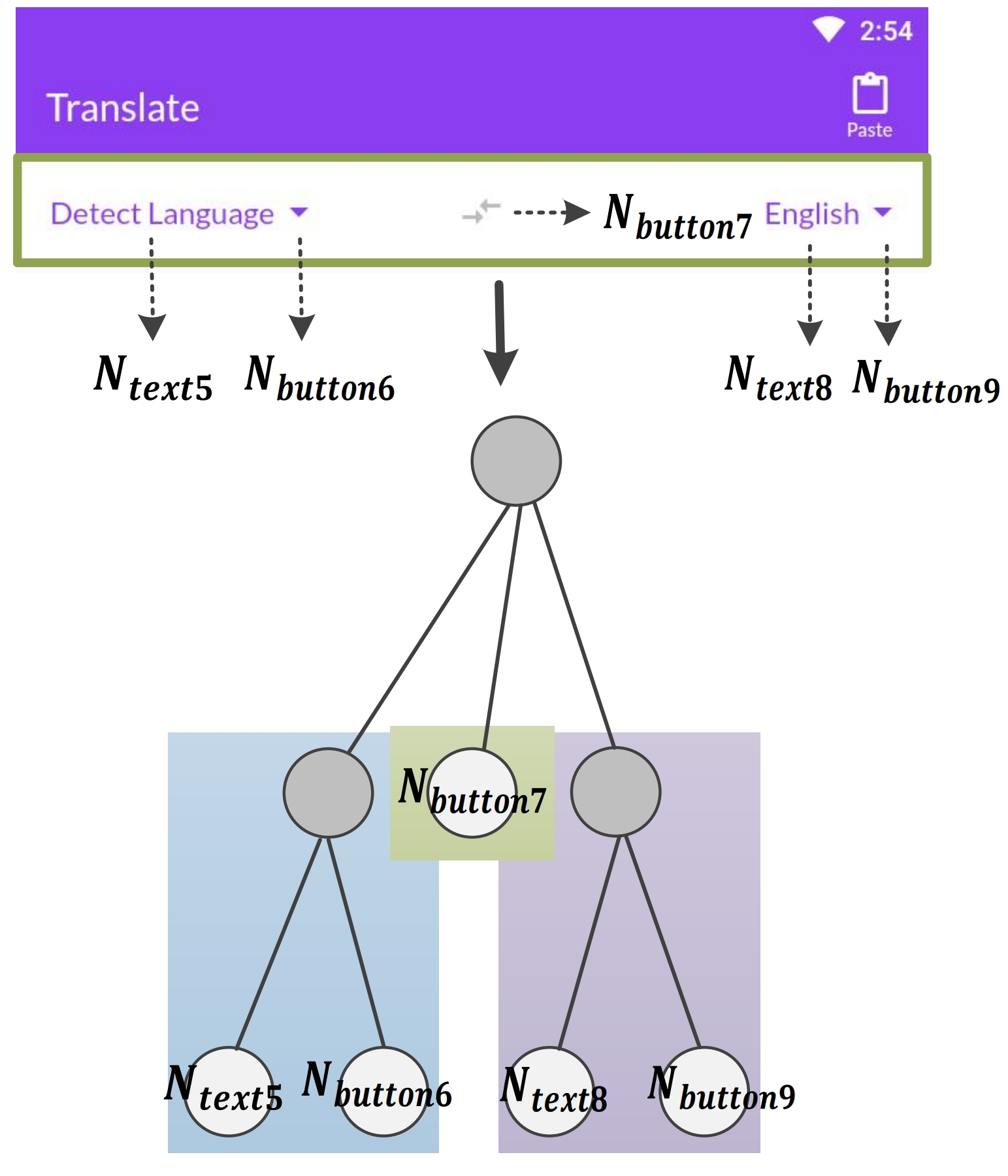}
\caption{An example of \emph{proximal-regions}.}
\label{fig: position}
\end{figure}

\subsubsection{Similarity strategy} Under this law, components within the GUI that share similar shapes and consistent view types are considered to be grouped. 
Therefore, within the \emph{initial-regions}, we further compared the view types and shapes of the components, grouped them accordingly, and designated them as \emph{similar-regions}.
As shown in algorithm~\ref{alg:group_similarity_shape}, we listed the reproducible procedures of this step and the associated pseudocode.
In the procedure of \textsc{Group\_Similar\_Regions}, if the view types are the same and the shapes are similar, we inferred that the navigation sequence for these component classes should be consecutive.
Notably, when the views are all \emph{$<$textview$>$}, we would not calculate their shape similarity but only group them based on the proximity strategy.
The view types of components can be directly obtained based on the view attributes in the GUI hierarchies.
However, the shapes of components, due to diverse design styles, lack an effective method for direct comparison.
Furthermore, we could not determine the shape of components solely based on their positional information, as the GUI hierarchy only provides information regarding the outer rectangular bounds of the components.
Thus, to obtain the shape of GUI components (the procedure of \textsc{Detectshape} in algorithm~\ref{alg:group_similarity_shape}), we employed the Canny edge detection algorithm~\cite{Canny1986ACA}~\footnote{A method in image processing for detecting edges in images while minimizing noise and accurately preserving the edges' location and shape.}.
The rationale for this selection is that the GUI components often consist of well-defined shapes, lines, and boundaries that need to be clearly distinguished for effective analysis.
The Canny edge detection algorithm could accurately identify and delineate edges and ensure that only significant edges are detected, reducing the likelihood of false positives.
In this process, we initially applied a lightweight tool named UIED~\cite{xie2020uied} for separating component images from GUI screenshots, and obtain the results.
Then, we adopted the Gaussian blur~\cite{Flusser2016RecognitionOI}~\footnote{A technique that reduces noise by weighting the average of each pixel in an image and its surrounding pixels to make the image smoother.} to the GUI component images, as indicated by the formula~\ref{for: gaussian}, to reduce noise in these images.
Specifically, GUIs often include complex backgrounds, gradients, and overlapping components, making shape extraction difficult without pre-processing. 
Also, certain GUI components may have sharp edges or high contrast, which might introduce noise and make it challenging to extract the shapes. 
Therefore, we employ the Gaussian Blur, a common pre-processing step that helps reduce noise in an image by smoothing out high-frequency components. 
In the context of a GUI, this reduces the impact of small, irrelevant details (e.g., pixel-level noise) that might otherwise interfere with shape detection. 
Besides, in GUIs, components often need to be detected in a consistent manner across different resolutions and styles. 
The Gaussian blur can help achieve a uniformity in detecting these shapes, regardless of small variations in the design of the GUI components.
\begin{equation}
\label{for: gaussian}
G(x, y) = \frac{1}{2\pi\sigma^2} e^{-\frac{x^2 + y^2}{2\sigma^2}}
\end{equation}
where $G(x, y)$ denotes the value of the Gaussian function at coordinates $(x, y)$, and $\sigma$ represents the standard deviation of the Gaussian distribution.

\begin{algorithm}
\footnotesize
\caption{Grouping Similar Regions}
\label{alg:group_similarity_shape}

\begin{algorithmic}[1]

\Procedure{Group\_Similar\_Regions}{initialRegions}
    \State \textbf{Input:} Set of initial regions $initialRegions$
    \State \textbf{Output:} Set of similar regions $similarRegions$
    
    \State Initialize $similarRegions \gets \emptyset$
    
    \For{each $region \in initialRegions$}
        \For{each pair of components $c_i, c_j \in region$}
            \If{ $c_i.viewType = c_j.viewType$}
               \If{ $c_i.viewType = $ $<$textview$>$ }
                  \State Group $c_i$ and $c_j$ into a $proximal-region$
               \Else
            	      \State $s_i \gets \textsc{DetectShape}(c_i)$
                  \State $s_j \gets \textsc{DetectShape}(c_j)$
   
                  \State $dist \gets \textsc{Hausdorff}(s_i, s_j)$
                    
                  \If{$dist \leq 0.1$} 
                        \State Group $c_i$ and $c_j$ into a $similar-region$
                  \EndIf
               \EndIf
            \EndIf
        \EndFor
        
        \State Add $similar-region$ to $similarRegions$
    \EndFor
    
    \State \textbf{return} $similarRegions$
\EndProcedure

\Procedure{DetectShape}{component images}
    \State \textbf{Input:} GUI component $component$ images
    \State \textbf{Output:} Detected shape $shape$
    
    \State $blurredImage \gets \textsc{GaussianBlur}(image)$ 
    
    \State $sobelImage \gets \textsc{SobelOperator}(blurredImage)$
    
    \State $edges \gets \textsc{Canny}(sobelImage)$ 
    
    \State $filtEdges \gets \textsc{NonMaxSuppression}(edges)$ 
    
    \State $finalEdges \gets \textsc{HysteresisThreshold}(filtEdges)$
    
    \State $shape \gets \textsc{ExtractShapeFromEdges}(finalEdges)$ 
    
    \State \textbf{return} $shape$
\EndProcedure

\Function{Hausdorff}{$A$, $B$}
    \State \textbf{Input:} Edge point sets of two shapes $A$, $B$
    \State \textbf{Output:} Hausdorff distance $H$
    
    \State $H \gets \max \left\{ \sup_{a \in A} \inf_{b \in B} d(a, b), \sup_{b \in B} \inf_{a \in A} d(b, a) \right\}$ 
    
    \State \textbf{return} $H$
\EndFunction

\end{algorithmic}
\end{algorithm}
Following the results of Gaussian blur, we proceeded to enhance the image's features by employing the Sobel operator~\cite{Han2020ResearchOE}.
It computes the gradient and direction for each pixel in the GUI component image, together with determining the edge strength of the image.
The rationale for using this operator lies in its efficiency in detecting gradients and highlighting edges in GUI images. 
GUI components typically exhibit strong contrast between interactive components (like buttons, images, and texts) and the background.
The boundary of GUI components can be determined using gradient information, where higher gradient values suggest a higher likelihood of the presence of boundaries in those components~\cite{Wu2021FedCGLC}. 
Moreover, we utilized gradient directions to ascertain the orientation of edges. 
The directional information aids in distinguishing between various types of edges within the GUI, contributing to a more nuanced understanding of the image.
Upon detecting potential edge areas, we further conducted non-maximum suppression~\cite{Hosang2017LearningNS} on the gradient values to eliminate redundant edge information in edge detection, retaining only the most representative edges. 
This involves evaluating each pixel in the gradient magnitude image and preserving only the local maxima in the direction of the gradient. 

In the subsequent stage, we adopted a hysteresis thresholding process~\cite{Yeh2017ThreeProngedCA} to find the edge information within the GUI components.
This process classifies pixels according to the strength of their gradients, distinguishing them into three principal categories: strong edges, weak edges, and non-edges.
Primarily, we established two threshold values, commonly referred to as the high threshold and low threshold, to discriminate pixels in the gradient image.
In our work, guided by parameters established in previous research~\cite{Sornam2016HysteresisTB}, we set the high threshold between 70\% and 90\%, while the low threshold ranges from 30\% to 50\%. 
If a pixel's gradient value surpasses the high threshold, it is designated as a strong edge, representing the most prominent boundaries or changes within the image.
For pixels whose gradient values fall between the low and high thresholds, they are deemed as weak edges, which often correspond to secondary edges or responses to noise~\cite{Eigenmann2017StructureLO}. 
Also, pixels with gradient values below the low threshold are categorized as non-edges, indicative of relatively smooth areas within the image lacking distinct edge structures. 
Then, we used connectivity analysis to connect strong edges and their adjacent weak edges.
Finally, we identified and linked edge segments that shared consistent orientations and were spatially contiguous, and further obtained the well-connected edge information of GUI components.

\begin{figure}
\centering
\includegraphics[width=6.5cm]{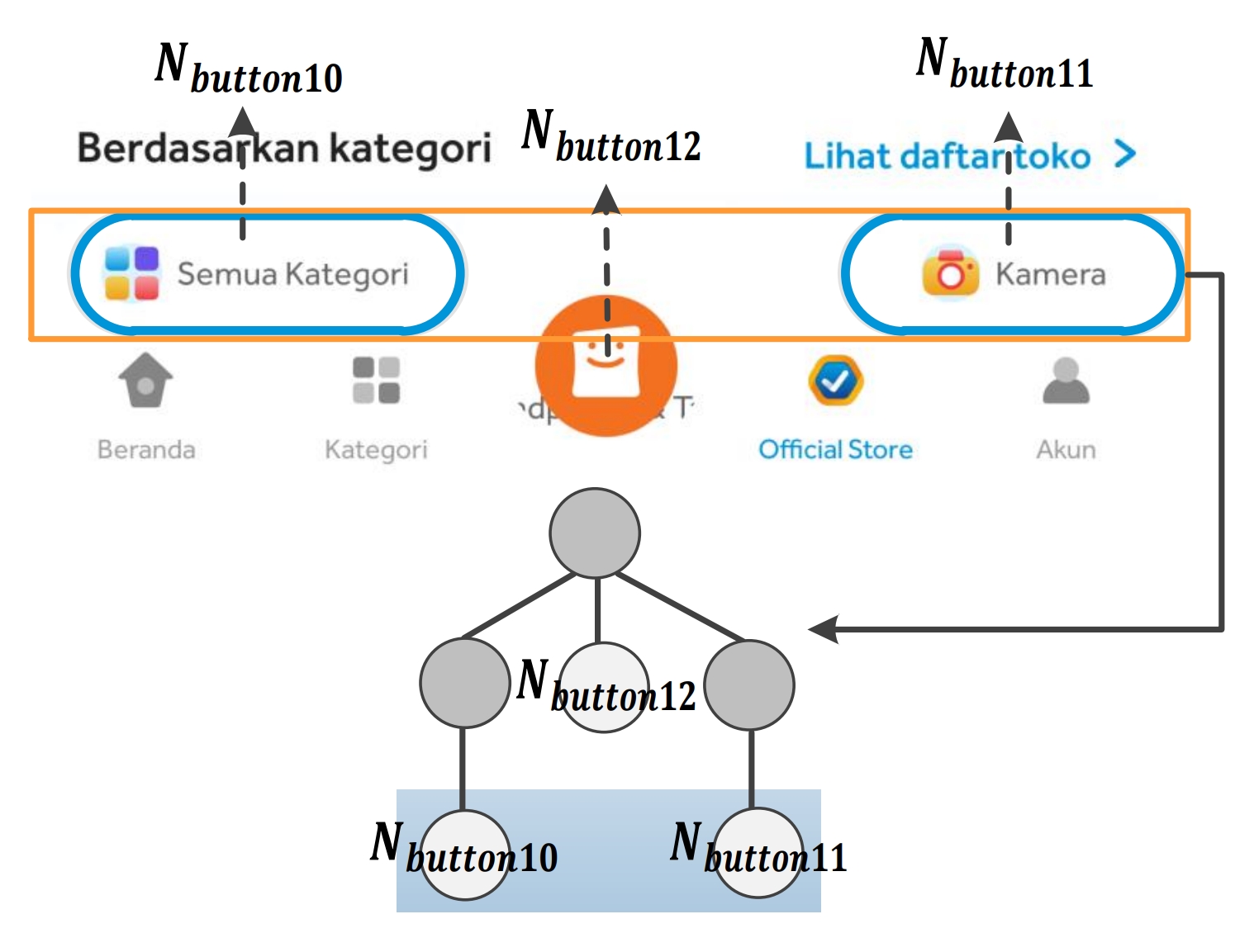}
\caption{An example of \emph{similar-regions}.}
\label{fig: similarity}
\end{figure}
\begin{figure*}
\centering
\includegraphics[width=17cm]{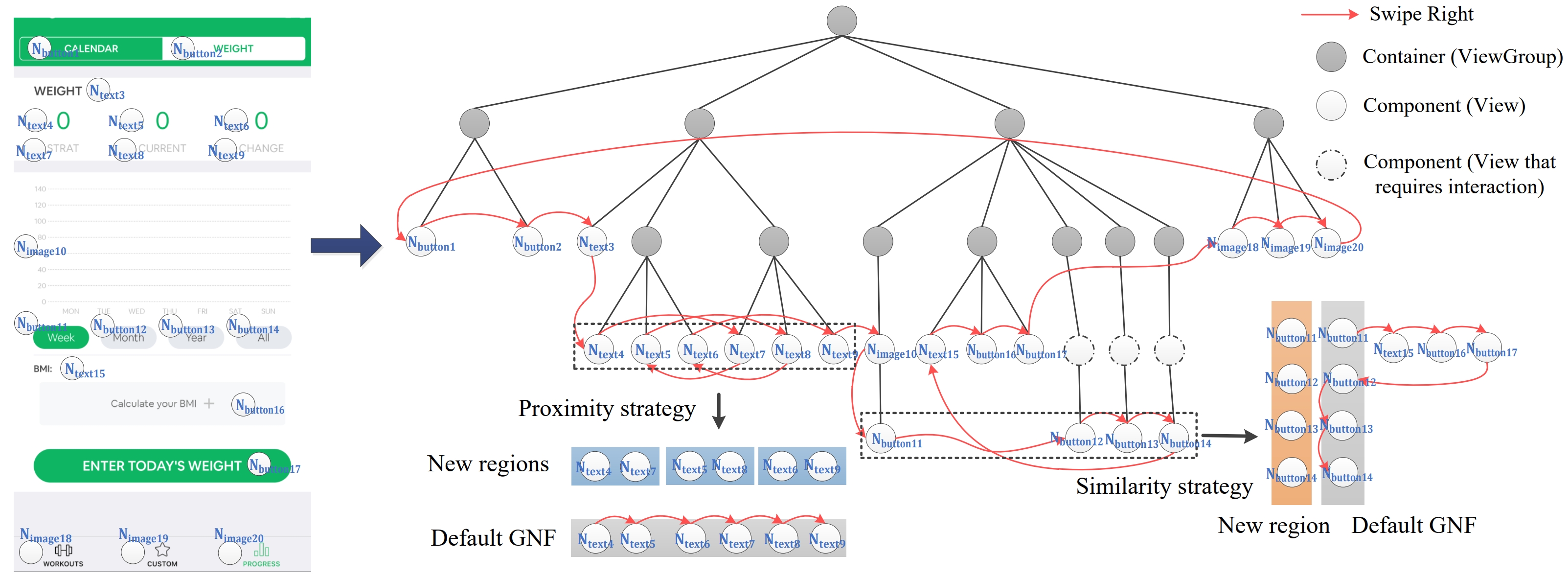}
\caption{A navigation process after redrawing. The red arrows indicate the navigation when the user swipes right.}
\label{fig: implementation}
\end{figure*}
At this point, we obtained the edge information of each GUI component and used it as a feature vector representation of its shape. 
Then, we computed the Hausdorff distance~\cite{Huttenlocher1993ComparingIU} between edge point sets of individual components within the same GUI container to assess the similarity of their edge shapes, as shown in the below formula~\ref{for: haus}.
\begin{equation}
\label{for: haus}
H(A, B) = \max \left\{ \sup_{a \in A} \inf_{b \in B} d(a, b), \sup_{b \in B} \inf_{a \in A} d(b, a) \right\}
\end{equation}
where $H(A, B)$ denotes the Hausdorff distance between two non-empty subsets $A$ and $B$, $d(a,b)$ represents the distance between elements $a$ and $b$ in the subsets.
This process can also be found in the function named \textsc{Hausdorff (A, B)} in algorithm~\ref{alg:group_similarity_shape}.
A smaller result indicates a higher degree of similarity between two edge shapes. 
Our decision to use the Hausdorff distance, as opposed to other similarity metrics, is based on the following two key factors: (1) \emph{Structural sensitivity}: the Hausdorff distance is adept at capturing the global structure and geometric characteristics of the images, which is crucial in the context of our study. While other similarity metrics like structural similarity index (SSIM)~\cite{nilsson2020understanding} or content-based measures are valuable, they are more focused on evaluating visual content, textures, or pixel-level differences. Our focus, however, is on the structural integrity of the GUI components, which makes Hausdorff distance a more appropriate choice for our specific needs~\cite{Kreveld2020BetweenSU}.
(2) \emph{Robustness:} this distance metric demonstrates resilience to small local variations, making it particularly effective in scenarios where noise or partial occlusions are present. 
Its robustness is essential for accurately matching complex shapes within our dataset~\cite{Karimi2019ReducingTH}.
Following the previous research~\cite{Chen2011ClusteringOT}, we define a threshold value of 0.1 to signify that the edges of the two components are considered similar. 
A specific instance is illustrated in Figure~\ref{fig: similarity}, where, following the constraints of the similarity rule, views of $N_{button10}$ and $N_{button11}$ with similar shapes ($H=0.015<0.1$) are assigned to the \emph{similar-region}.
Without this region constraint, the screen reader would navigate to the bottom navigation component $N_{button12}$ after accessing component $N_{button10}$, instead of consecutively reading $N_{button11}$, which shares a similar shape.

\subsection{GNF redrawing process}

Considering the characteristics of visually impaired users' perceptions, which entails continuous reading of components with proximity and similar positions, we have grouped GUI components into two distinct regions: \emph{proximal-regions} and \emph{similar-regions}.
In this section, we redrew the GNF based on these regions combined with the GUI hierarchy.
Figure~\ref{fig: implementation} shows an example of our redrawn GNF.
On the left side of this figure is the GUI, while on the right side is the GUI view tree along with the GNF is highlighted with red arrows.
The blue and orange areas in this figure respectively exemplify the \emph{proximal-region} and \emph{similar-region} we have marked.

When navigating the GUI, our method adheres to the following two rules: 
\begin{enumerate}
\item \emph{Rule 1:} It dictates navigation within the GUI hierarchy by traversing component nodes proceed with depth-first search (DFS).
\item \emph{Rule 2:} It stipulates that when the screen reader focuses resides within the defined \emph{proximal-regions} and \emph{similar-regions}, it must navigate through all components within the region before progressing to the next region.
\end{enumerate}

Taking the example illustrated in Figure~\ref{fig: implementation}, when the screen reader navigates this GUI, it first visits the component $N_{button1}$ according to \emph{Rule 1}, which is the component node positioned at the leftmost in the GUI view tree.
Continuing to follow \emph{Rule 1}, the screen reader sequentially navigates towards components $N_{button2}$, $N_{text3}$, and $N_{text4}$ in the view tree.
At this point, as $N_{text4}$ and $N_{text7}$ are assigned to the same \emph{proximal-region}, according to \emph{Rule 2}, $N_{text7}$ should be navigated to after $N_{text4}$.
Following this, navigation proceeds to component $N_{text5}$ to the right of $N_{text4}$, and as $N_{text5}$ belongs to the same \emph{proximal-region} as $N_{text8}$, and $N_{text8}$ is navigated after $N_{text5}$. 
Similarly, the component navigated to after $N_{text6}$ is $N_{text9}$.
The above navigation process of these six components ($N_{text4}$ $\rightarrow$ $N_{text7}$ $\rightarrow$ $N_{text5}$ $\rightarrow$ $N_{text8}$ $\rightarrow$ $N_{text6}$ $\rightarrow$ $N_{text9}$) demonstrates the advantage of our partitioned \emph{proximal-regions}.
In this example, these six components should present a navigation that matches the number and the textual cues one by one.
If the screen reader navigates according to the default order, the GNF would be $N_{text4}$ $\rightarrow$ $N_{text5}$ $\rightarrow$ $N_{text6}$ $\rightarrow$ $N_{text7}$ $\rightarrow$ $N_{text8}$ $\rightarrow$ $N_{text9}$ (the gray area on the bottom side of Figure~\ref{fig: implementation}).
This navigation method tends to disjoint numerical information from its related textual cues, leading to mismatched information. 

Following the navigation above, we now have navigated the component $N_{text9}$.
After this, according to \emph{Rule 1}, the screen reader moves to the region where $N_{image10}$ is located and then proceeds to component $N_{button11}$.
At this point, components $N_{button11}$, $N_{button12}$, $N_{button13}$, and $N_{button14}$ are assigned to the same \emph{similar-region}.
Therefore, following \emph{Rule 2}, these four components need to be navigated consecutively, with the sequence being $N_{button11}$ $\rightarrow$ $N_{button12}$ $\rightarrow$ $N_{button13}$ $\rightarrow$ $N_{button14}$.
In this process, the \emph{similar-regions} we grouped possess advantages.
The default navigation of these four components is shown in the gray area on the right side of Figure~\ref{fig: implementation}, where after reading $N_{button11}$, the screen reader proceeds to $N_{text15}$, $N_{button16}$, and $N_{button17}$ before reading $N_{button12}$.
This sequence disrupts the flow of continuous information for visually impaired users.
Conversely, our redrawn GNF preserves this continuity well.
Then, according to DFS, the navigation continues downward on the GUI, and the screen reader sequentially navigates components $N_{text15}$, $N_{button16}$, and $N_{button17}$. 
Finally, the screen reader reaches the bottom navigation bar, and according to \emph{Rule 1}, it navigates components $N_{image18}$, $N_{image19}$, and $N_{image20}$ in sequence.
It is important to note that after reaching component $N_{image20}$, users who continue to swipe right will return to the component $N_{button1}$.

\begin{figure}
\centering
\includegraphics[width=6.5cm]{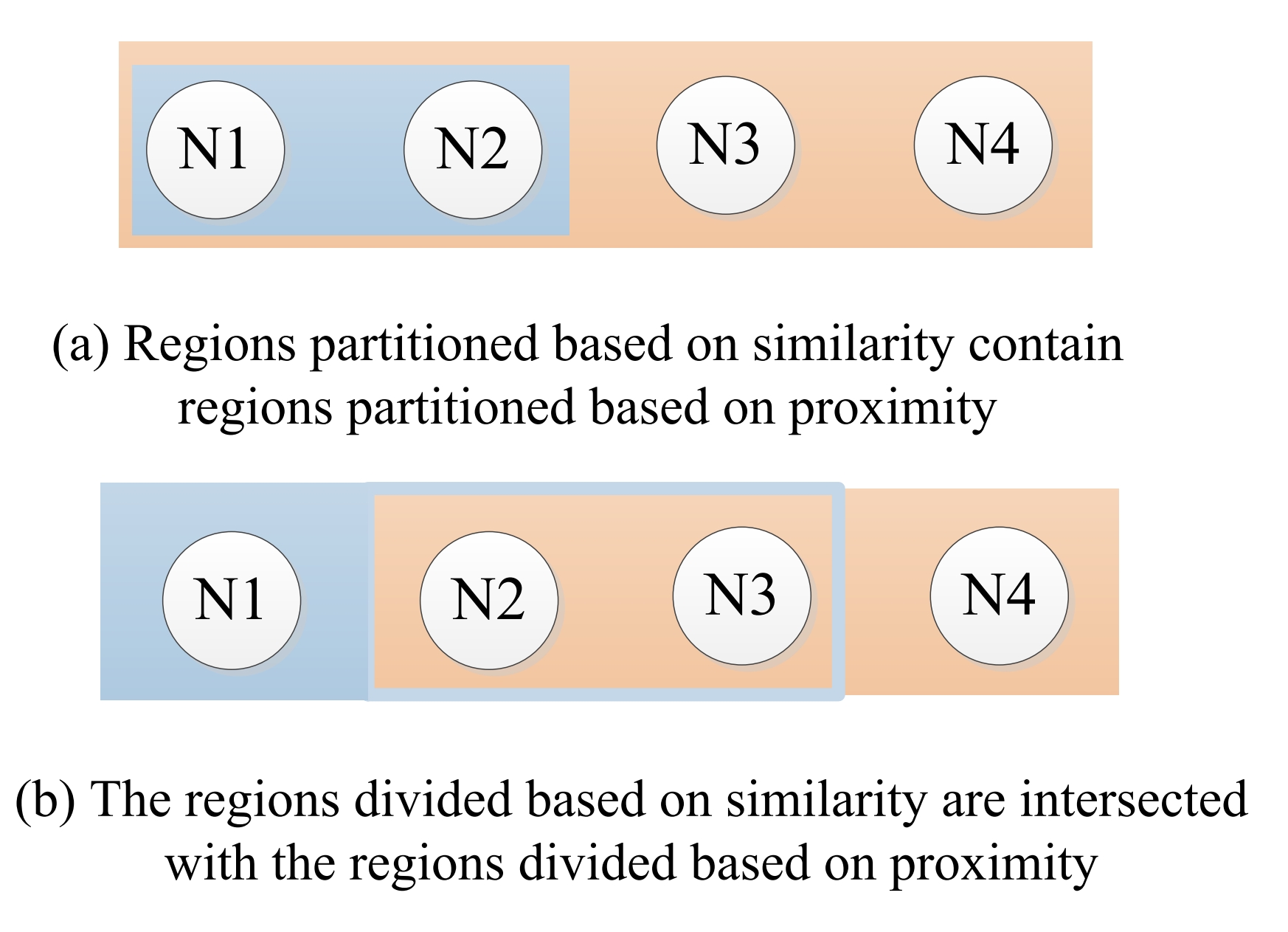}
\caption{An example where the divided regions include or intersect.}
\label{fig: intersect}
\end{figure}
Furthermore, in the illustrated example depicted in Figure~\ref{fig: implementation}, some components require user interactions to be displayed, notably observed in component $N_{button12}$ where a double-click is necessary to trigger a switch. 
However, as this GUI does not directly present its information, we excluded it from this navigation process for a while.
We also found that the regions divided by the two laws may include or intersect.
As shown in Figure~\ref{fig: intersect}, this may manifest as one region containing the other (Figure~\ref{fig: intersect} (a)), or two regions intersecting but not containing (Figure~\ref{fig: intersect} (b)).
In this case, our method might become uncertain about whether the elements within that region are governed by the law of proximity or similarity, potentially leading to repeated reading of components within the overlapping area. 
Therefore, we configured the navigation to proceed continuously.
For instance, in Figure~\ref{fig: intersect} (b), the navigation would consecutively read the elements within the two overlapping views as follows: $N_{button1}$ $\rightarrow$ $N_{button2}$ $\rightarrow$ $N_{button3}$ $\rightarrow$ $N_{button4}$.

\subsection{Implementation}
We have implemented the approach mentioned above into a tool, also named RGNF, to redraw the GNFs for mobile apps.
The preprocess of GUI hierarchies is presented in the file named \emph{UI\_Navigation\_Process}.
The implementations of redrawing GNFs are written in Kotlin and compiled to run on JavaScript. 
The \emph{AccessibilityHelper.kt} file invokes the \emph{onAccessibilityEvent()} method and \emph{AccessibilityNodeinfo()} method of the Accessibility Service to extract the GUI hierarchy.
The GNF is then obtained in the \emph{MainActivity.kt} file and redrawn according to the Gestalt psychological model. 
Further, this project is developed using Kotlin 1.8.0 and executed in the Eclipse IDE, with the graphics card required being NVIDIA GTX 2080Ti. 
It is important to note that, to ensure the proper invocation of the Accessibility Service, users need to install the Android SDK locally and complete the relevant configurations.

Using the aforementioned scripts and configuration, we apply the redrawn GNF to the app through the following process. 
First, we identify each component in the GNF by its view type and resource\_id, and set the navigation order based on our redrawing results. 
Second, navigation begins with the first component, and we search for the subsequent component according to the established order. 
By setting the \emph{nextFocusRight()} attribute of the first component to the resource\_id of the next component, we enable seamless navigation between these two components. 
Finally, we repeat this process for all components in the GNF to establish the desired navigation sequence.

\section{Evaluation of RGNF}\label{sec: experiment}

In this section, we embarked on the evaluation of our proposed method. 
The objective of this evaluation is to assess the effectiveness of our approach in redrawing the GNF, and ensure its intended purposes are achieved.
We structured our evaluation around two key points: effectiveness (RQ1) evaluated by experiments and usefulness (RQ2) evaluated by a user study, as these were paramount in determining the overall performance and user experience.
The two research questions we addressed were presented below.
\begin{itemize}
\item \textbf{RQ1 (Effectiveness):} Can our redrawn GNF effectively facilitate visually impaired users to access GUI components?
\item \textbf{RQ2 (Usefulness):} What is the actual evaluation of the redrawn GNF by visually impaired users regarding our approach?
\end{itemize}

In RQ1, we first calculated the similarity between our redrawn GNF and the true set, and compared it with the similarity between the default GNF and the true set.
Further, leveraging the tool proposed by Alotaibi et al.~\cite{Alotaibi2022AutomatedDO}, we assessed the reachability of each GUI component when we used the redrawn GNF and compared it with the default navigation provided by Talkback. 
We also carried out an ablation study on the filtering process we applied to determine its impact on the method's effectiveness.
As for RQ2, we focused on the actual user experience of visually impaired users. 
In this user study, we asked visually impaired users to navigate the GUIs under the default and redrawn GNFs to recorded their rating results, and analyzed their feedback on which GNF was more useful.

\subsection{RQ1: Evaluation of Effectiveness by Experiments}

\subsubsection{Data collection}\label{sub: data}

Our dataset consists of 25 popular apps (downloads$\geq$5 million) from the recommended list on Google Play and 5 apps with fewer downloads ($\leq$10K).
There are two reasons for selecting apps in this way.
One is that it is difficult for us to test all of the more than 2 million apps hosted by Google Play.
Thus, following the previous works~\cite{Liu2020OwlES, Bajammal2021SemanticWA, Li2022PushButtonSO}, we selected a small subset to conduct experiments and illustrated the effectiveness of our method.
The other is that the comprehensive evaluation of popular apps and unpopular apps can better evaluate the universality of our method.
The names of these apps are listed in the first column of Table~\ref{tab: effectiveness}. 
Among them, the top 25 apps in this table are from the recommended list, while the remaining 5 apps are less downloaded.
Besides, the number of GUIs we obtained in each app is listed in the parentheses of the first column. 
These GUIs were manually captured, and we excluded GUIs with little information and simple structure, such as login pages, as visually impaired users could access information on such pages regardless of the navigation strategy employed.
Conversely, we retained GUIs with complex designs, such as homepages.

Furthermore, our experimental ground truth was created by visually impaired users based on the 30 selected apps. 
This dataset encompassed the expected navigation sequence of all components as perceived by visually impaired users in their cognitive understanding.
To this end, we first recruited 10 visually impaired users (5 male and 5 female) from the School of Special Education. 
It was worth noting that these users were different from the users we investigated in Section~\ref{sub: psy}, to ensure that there was no bias in the construction of the true set.
Before the marking process, participants underwent a training session to acquaint themselves with the specific tasks.
They were introduced to the study's objectives and guidelines for marking navigation sequences.
These 30 apps were then distributed among 10 participants, with each app assigned to two participants for annotation, mitigating the impact of human subjectivity on the results.
After that, participants were instructed to interact with the apps they were responsible for. 
As they navigate through the GUI components, they were asked to verbally articulate their expected sequence of interactions. 
This included stating which component they expected to encounter next, and the reasons behind their expectations.
Finally, the participants' verbal descriptions were recorded and transcribed.
The first three authors carefully reviewed the transcriptions and logged to validate the marked navigation sequences. 
Any discrepancies or ambiguities were clarified with the participants in time.

Throughout this process, we maintained the ethical standards for all participants. 
We obtained informed consent from all participants, ensuring they understood the purpose and scope of the study. 
Further, we upheld privacy and anonymity, and prepared data security measures to protect participants' personal information.

\subsubsection{Research settings}

To answer this research question, we conducted three experiments.
The first one involved investigating the similarity between the redrawn GNFs and the ground truth, as well as the evaluation between the default GNFs (baseline) and the ground truth.
The baseline refers to the app's default navigation flow provided by Talkback, while the ground truth represents the navigation flow identified by visually impaired users based on their expected navigation patterns.
To this end, we employed the Euclidean distance~\cite{Deza2014EncyclopediaOD} to compute the similarity.
Specifically, the components in the true set were arranged one by one according to the order of their navigation flow, with the first node assigned a value of 1, the second node assigned a value of 2, and so forth.
The components in the redrawn GNFs and the baseline were also arranged in sequence according to the order in the ground truth, but their assignment of values was based on their respective navigation sequences.
Subsequently, we calculated the similarity $S$ according to the following formula~\ref{for: Euclidean}.
\begin{equation}
\label{for: Euclidean}
S(x,y) = \frac{1}{1+\sqrt{\sum{(x_i-y_i)^2}}}
\end{equation}
where $x$ represents the ground truth, and $y$ is the result of the redrawn GNFs or the default GNFs.
$x_i$ denotes the specific value assigned to each component in ground truth, and $y_i$ refers to the values assigned to redrawn GNFs or default GNFs.
The value of $S$ is in the range $(0,1]$, with the closer to 1 indicating the more similar the two are.

Furthermore, we utilized the tool proposed by Alotaibi et al.~\cite{Alotaibi2022AutomatedDO} to assess the reachability of our redrawn GNFs and default GNFs.
The output consists of two values: one indicating the number of components reachable by swipe navigation starting from the top-left corner of the GUI ($N$), and the other indicating the number of components reachable by touch browsing and swipe navigation ($M$).
If $N=M$, it indicates that the components that visually impaired users can access through swipe navigating and touch browsing are consistent.
However, if $N$ is lower than $M$, it suggests that there are components that cannot be accessed while swiping navigation.
The rationale for assessing reachability lies in the fact that the basis of navigation for visually impaired users within GUIs relies on acquiring information as comprehensively as possible. 
If there are too many unreachable components in a navigation process, it severely hampers visually impaired users' ability to access information.

The third experiment surrounding this research question was an ablation study, aiming to investigate the impact of the filtering process within our method on its effectiveness. 
To achieve this purpose, we kept the invisible and unfocusable views in GNFs, denoted the new method as RGNF\_\emph{w/o\_VF}. 
Subsequently, we conducted assessments of sequence similarity and reachability (the aforementioned two experiments) on this method.
Finally, we qualitatively analyzed the differences between RGNF and RGNF\_\emph{w/o\_VF}.

\begin{figure*}
\centering
\includegraphics[width=16.5cm]{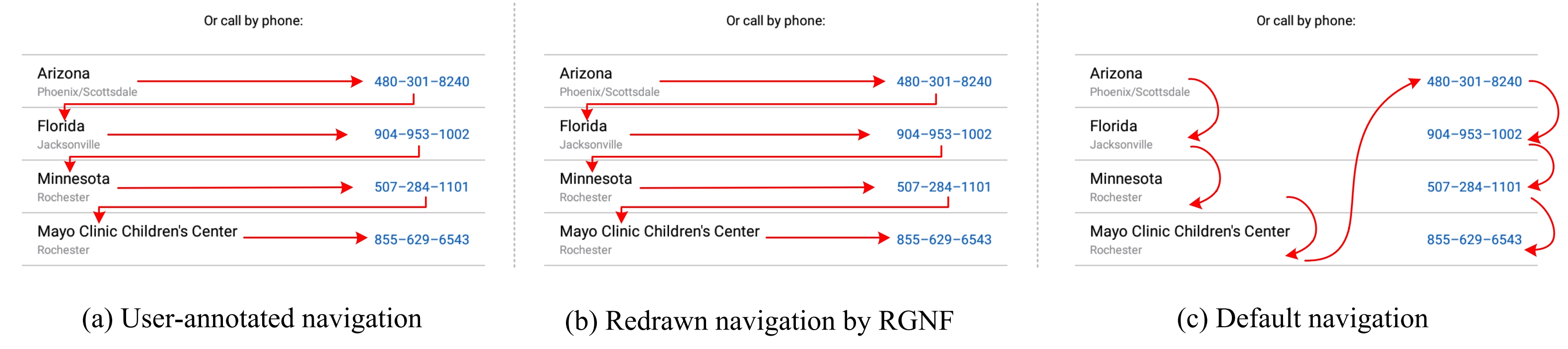}
\caption{A practical GUI navigation example for swiping to the right, comprising the user-annotated ground truth, the results redrawn by our method, and the default navigation sequence.}
\label{fig: experiment}
\end{figure*}

\begin{table}[!t]\scriptsize
\renewcommand{\arraystretch}{1.4}
\tabcolsep=0.1cm
\caption{Effectiveness of RGNF.} \label{tab: effectiveness}
\begin{center}
\begin{tabular}{lcccc}
\hline
\textbf{Apps} \textcolor{gray}{(GUI)}& \textbf{S\_Ours} & \textbf{S\_baseline} & \textbf{R\_ours} \textcolor{gray}{(\%)} & \textbf{R\_baseline} \textcolor{gray}{(\%)} \cr
\hline

\rowcolor{gray!3}1. Sky Map \textcolor{gray}{(5)} & 0.924 & 0.611 & 65/74 \textcolor{gray}{(87.84)} & 55/74 \textcolor{gray}{(74.32)}\cr
\rowcolor{gray!8}2. Oxford \textcolor{gray}{(4)} & 0.92 & 0.638 & 42/48 \textcolor{gray}{(87.5)} & 34/48 \textcolor{gray}{(70.83)}\cr
\rowcolor{gray!3}3. OverDrive \textcolor{gray}{(6)} & 0.941 & 0.641 & 71/79 \textcolor{gray}{(89.87)} & 62/79 \textcolor{gray}{(78.48)}\cr
\rowcolor{gray!8}4. NewPipe \textcolor{gray}{(5)} & 0.927 & 0.648 & 36/41 \textcolor{gray}{(87.8)} & 29/41 \textcolor{gray}{(70.73)}\cr
\rowcolor{gray!3}5. Starbucks \textcolor{gray}{(8)} & 0.906 & 0.622 & 119/124 \textcolor{gray}{(95.97)} & 101/124 \textcolor{gray}{(81.45)}\cr
\rowcolor{gray!8}6. OpenTable \textcolor{gray}{(6)} & 0.921 & 0.628 & 72/78 \textcolor{gray}{(92.31)} & 59/78 \textcolor{gray}{(75.64)}\cr
\rowcolor{gray!3}7. Snapseed \textcolor{gray}{(4)} & 0.935 & 0.642 & 22/25 \textcolor{gray}{(88.00)} & 18/25 \textcolor{gray}{(72.00)}\cr
\rowcolor{gray!8}8. Wise \textcolor{gray}{(9)} & 0.93 & 0.609 & 272/289 \textcolor{gray}{(94.12)} & 233/289 \textcolor{gray}{(80.62)}\cr
\rowcolor{gray!3}9. Grab \textcolor{gray}{(11)} & 0.919 & 0.635 & 266/274 \textcolor{gray}{(97.08)} & 224/274 \textcolor{gray}{(81.75)}\cr
\rowcolor{gray!8}10. Andev \textcolor{gray}{(6)} & 0.913 & 0.613 & 65/74 \textcolor{gray}{(87.84)} & 54/74 \textcolor{gray}{(72.97)}\cr
\rowcolor{gray!3}11. MyChart \textcolor{gray}{(6)} & 0.909 & 0.625 & 57/64 \textcolor{gray}{(89.06)} & 46/64 \textcolor{gray}{(71.87)}\cr
\rowcolor{gray!8}12. Indeed Jobs \textcolor{gray}{(6)} & 0.931 & 0.619 & 48/53 \textcolor{gray}{(90.56)} & 38/53 \textcolor{gray}{(71.69)}\cr
\rowcolor{gray!3}13. ESPN \textcolor{gray}{(10)} & 0.943 & 0.614 & 102/114 \textcolor{gray}{(89.47)} & 88/114 \textcolor{gray}{(77.19)}\cr
\rowcolor{gray!8}14. NBC News \textcolor{gray}{(4)} & 0.908 & 0.646 & 37/42 \textcolor{gray}{(88.09)} & 30/42 \textcolor{gray}{(71.43)}\cr
\rowcolor{gray!3}15. Telegram \textcolor{gray}{(9)} & 0.933 & 0.615 & 93/101 \textcolor{gray}{(92.08)} & 75/101 \textcolor{gray}{(74.26)}\cr
\rowcolor{gray!8}16. WhatsApp \textcolor{gray}{(8)} & 0.911 & 0.623 & 86/97 \textcolor{gray}{(88.66)} & 70/97 \textcolor{gray}{(72.16)}\cr
\rowcolor{gray!3}17. Google Chat \textcolor{gray}{(7)} & 0.923 & 0.639 & 59/66 \textcolor{gray}{(89.39)} & 48/66 \textcolor{gray}{(72.73)}\cr
\rowcolor{gray!8}18. Google LLC \textcolor{gray}{(7)} & 0.932 & 0.633 & 113/124 \textcolor{gray}{(91.12)} & 87/124 \textcolor{gray}{(70.16)}\cr
\rowcolor{gray!3}19. TED \textcolor{gray}{(7)} & 0.902 & 0.624 & 82/89 \textcolor{gray}{(92.13)} & 72/89 \textcolor{gray}{(80.89)}\cr
\rowcolor{gray!8}20. LinkedIn \textcolor{gray}{(6)} & 0.926 & 0.631 & 68/72 \textcolor{gray}{(94.44)} & 51/72 \textcolor{gray}{(70.83)}\cr
\rowcolor{gray!3}21. Waze \textcolor{gray}{(7)} & 0.929 & 0.645 & 71/80 \textcolor{gray}{(88.75)} & 58/80 \textcolor{gray}{(72.50)}\cr
\rowcolor{gray!8}22. Printerest \textcolor{gray}{(8)} & 0.934 & 0.632 & 85/92 \textcolor{gray}{(92.39)} & 69/92 \textcolor{gray}{(75.00)}\cr
\rowcolor{gray!3}23. Wechat \textcolor{gray}{(8)} & 0.917 & 0.630 & 94/101 \textcolor{gray}{(93.07)} & 78/101 \textcolor{gray}{(77.23)}\cr
\rowcolor{gray!8}24. YouTube \textcolor{gray}{(11)} & 0.937 & 0.627 & 162/172 \textcolor{gray}{(94.19)} & 140/172 \textcolor{gray}{(81.39)}\cr
\rowcolor{gray!3}25. Twitter \textcolor{gray}{(11)} & 0.918 & 0.634 & 184/192 \textcolor{gray}{(95.83)} & 154/192 \textcolor{gray}{(80.21)}\cr
\rowcolor{gray!8}26. Maverick \textcolor{gray}{(8)} & 0.875 & 0.537 & 49/60 \textcolor{gray}{(81.67)} & 35/60 \textcolor{gray}{(58.33)}\cr
\rowcolor{gray!3}27. Timber \textcolor{gray}{(5)} & 0.892 & 0.619 & 41/47 \textcolor{gray}{(87.23)} & 30/47 \textcolor{gray}{(63.82)}\cr
\rowcolor{gray!8}28. V2ex \textcolor{gray}{(7)} & 0.925 & 0.610 & 49/57 \textcolor{gray}{(85.96)} & 39/57 \textcolor{gray}{(68.42)}\cr
\rowcolor{gray!3}29. Auro \textcolor{gray}{(6)} & 0.938 & 0.620 & 87/94 \textcolor{gray}{(92.55)} & 74/94 \textcolor{gray}{(78.72)} \cr
\rowcolor{gray!8}30. Leisure \textcolor{gray}{(7)} & 0.922 & 0.618 & 102/108 \textcolor{gray}{(94.44)} & 90/108 \textcolor{gray}{(83.33)}\cr

\hline
\rowcolor{gray!3}Average & 0.921 & 0.624 & 90.31 & 74.35\cr
\hline
\end{tabular}
\end{center}
\end{table}

\subsubsection{Results and analysis} Table~\ref{tab: effectiveness} presents the results of evaluating our redrawn GNFs and the baseline for these 30 apps.
The first column is the name of the selected apps and the number of GUIs.
The second column presents the sequence similarity values ($S$) between our redrawn GNFs and the ground truths.
The third column displays the similarity between the default GNFs and the ground truths. 
In the fourth and fifth columns, we calculate the average reachability ratings of the redrawn GNFs and the baseline GNFs, as assessed by Alotaibi et al.'s tool. 
The results for each cell are presented as $N/M=P$ and $P$ is the ratio.
The last row is the average result of the effectiveness.

Compared to the true set we constructed, our redrawn GNF outperformed the default baseline.
One intuitive result is that the average sequence similarity value of our method is $S=0.921$, while the result of the baseline is $S=0.624$.
We further analyzed the reasons for this and found that the bias presented in the baseline sequence was mainly concentrated in the union form. 
The default navigation did't adjust the navigation flow to account for how the components are laid out in either view, they only followed the navigation in column-first order.
Figure~\ref{fig: experiment} illustrates a specific example, including the user-annotated ground truth (a), the results after redrawing (b), and the default navigation sequence (c). 
It was noticed that, with the default navigation (c), the coherence between the product names and prices was disrupted, making it challenging for users to locate or accurately perceive their corresponding relationships. 
In contrast, the redrawn GNF using our method (b) aligned consistently with the result represented in the user-annotated result (a).
Another situation worth analyzing was why the similarity exhibited by the popular apps we selected is so low. 
This is primarily because we found that although these apps are popular, the navigation flow they follow under Talkback adheres strictly to a top-to-bottom, left-to-right sequence, without modification or alternative solutions. 
This essentially reflects the developers' oversight of issues related to component navigation, further motivating our research. 
Certainly, we also acknowledge that these popular apps do not completely neglect their accessibility features, but their focus is primarily on the static layout of the GUI, such as component size, spacing, color, and content-descriptions. 
Therefore, our proposed approach offers valuable contributions: first, by highlighting the importance of addressing issues in the component navigation flow, and second, by providing developers with a navigation flow that aligns better with the expectations of visually impaired users.

As observed, the GNFs redrawn by our method exhibited an average reachability of 90.31\%, indicating that the components reachable through touch browsing in our redrawn GNFs were also well reachable through swipe navigation.
In contrast, when employing the default navigation flow, only 74.35\% of the components were reachable.
Qualitatively, the reasons why components are not reachable under the default navigation flow can be attributed to two aspects.
One aspect pertains to the inherent layout structure of the GUI, where certain components are nested within complex hierarchies or obscured by overlapping elements, making them difficult for screen readers to identify and access effectively. 
The other aspect is related to the ordering strategy, as screen readers may not prioritize these components correctly in the default sequence, resulting in users missing out on vital information or features due to the inefficient navigation flow~\cite{Rodrigues2015GettingST}.
As for the GNFs redrawn through our method, they extracted components accessible to visually impaired users from the view hierarchy, ensuring screen readers' effective reading and demonstrating improved reachability.

\begin{figure}
\centering
\includegraphics[width=6.5cm]{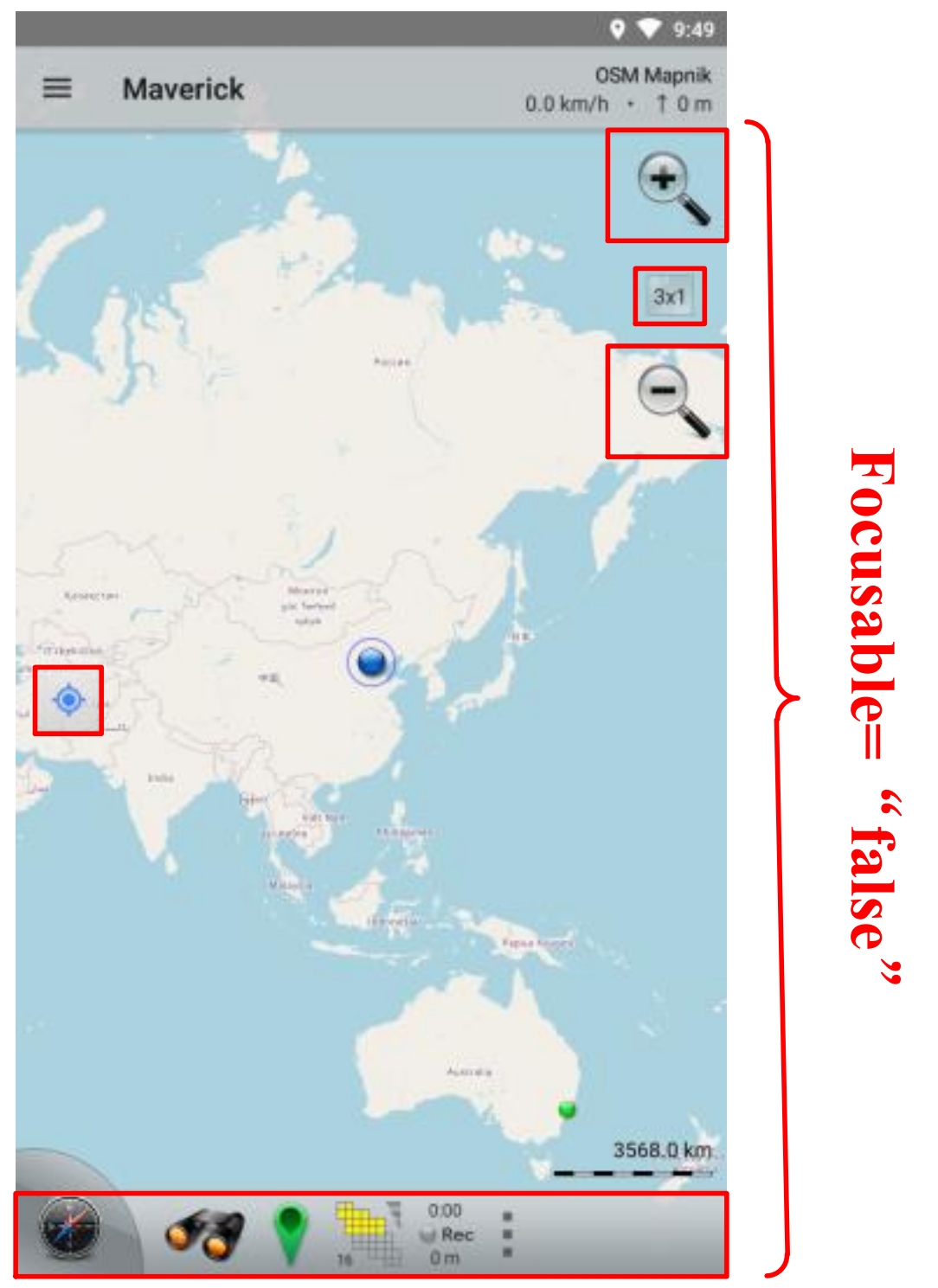}
\caption{The failure case when redrawing GNF.}
\label{fig: failure}
\end{figure}
In addition, we found that our method can be also effectively applied to the less downloaded apps, as evidenced by the improved similarity (increased by 0.3096 on average) and reachability (increased by 17.85\% on average) scores for the last five apps in Table~\ref{tab: effectiveness}. 
However, we noted that after redrawing the GNF of \emph{``Maverick''}, the similarity score was 0.875 and the reachability is 81.67\%, which was relatively low compared to the other apps. 
Such a scenario was mainly because this app has a GUI that RGNF cannot adjust, representing a failure case in our approach.
Specifically, Figure~\ref{fig: failure} presents this case.
Although this app presents numerous components, those encircled with red borders are incapable of receiving focus (\emph{focusable=``false''}).
This means that screen readers cannot access these components, leading to a failure in producing a redrawn GNF.
This failure can be attributed to developers mistakenly designating crucial functional components as unfocusable or, in many cases, neglecting the accessibility considerations for these components during the app's design phase. 
A prospective solution might involve detecting and rectifying the focus issues before redrawing the GNF, ensuring that screen readers can access these components.

We now delved into a detailed analysis of how our filtering process influences the effectiveness, as indicated by the results presented in the ablation study.
Figure~\ref{fig: ablation} shows the results of RGNF\_\emph{w/o\_VF} compared with RGNF in terms of sequence similarity and reachability.
As we can see, when filtering process (RGNF\_\emph{w/o\_VF}), there is a noticeable decline in these values. 
In the absence of filtering, the similarity values are lower ($S$(RGNF\_\emph{w/o\_VF})=0.537) than ours.
This reduction is primarily attributed to the presence of numerous invisible GUI components, causing screen readers to focus on them during navigation.
Thus, it would output extra information that makes the navigation flow different from what the users have annotated.
Furthermore, the reachability ($D$) also provided insights into the effectiveness of our approach. 
The average reachability of RGNF\_\emph{w/o\_VF} drops to 75.72\%, which is lower than the 90.31\% achieved by the RGNF. 
The rationale behind this variation is the retention of unfocusable components.
Users cannot access such components through swipe navigation, they are only accessible via touch exploration.
Therefore, retaining these components increases the denominator ($M$) when computing the score of reachability, consequently leading to decreased results.
In practice, even though these components may be reachable via touch, screen readers do not output their contents, rendering them invalid for visually impaired users.
This ablation study implies that our filtering process enhances accessibility and fosters a more cohesive and user-friendly navigation flow, providing better guidance for developers to address accessibility issues that arise during the navigation process.

\begin{figure}
\centering
\includegraphics[width=8.7cm]{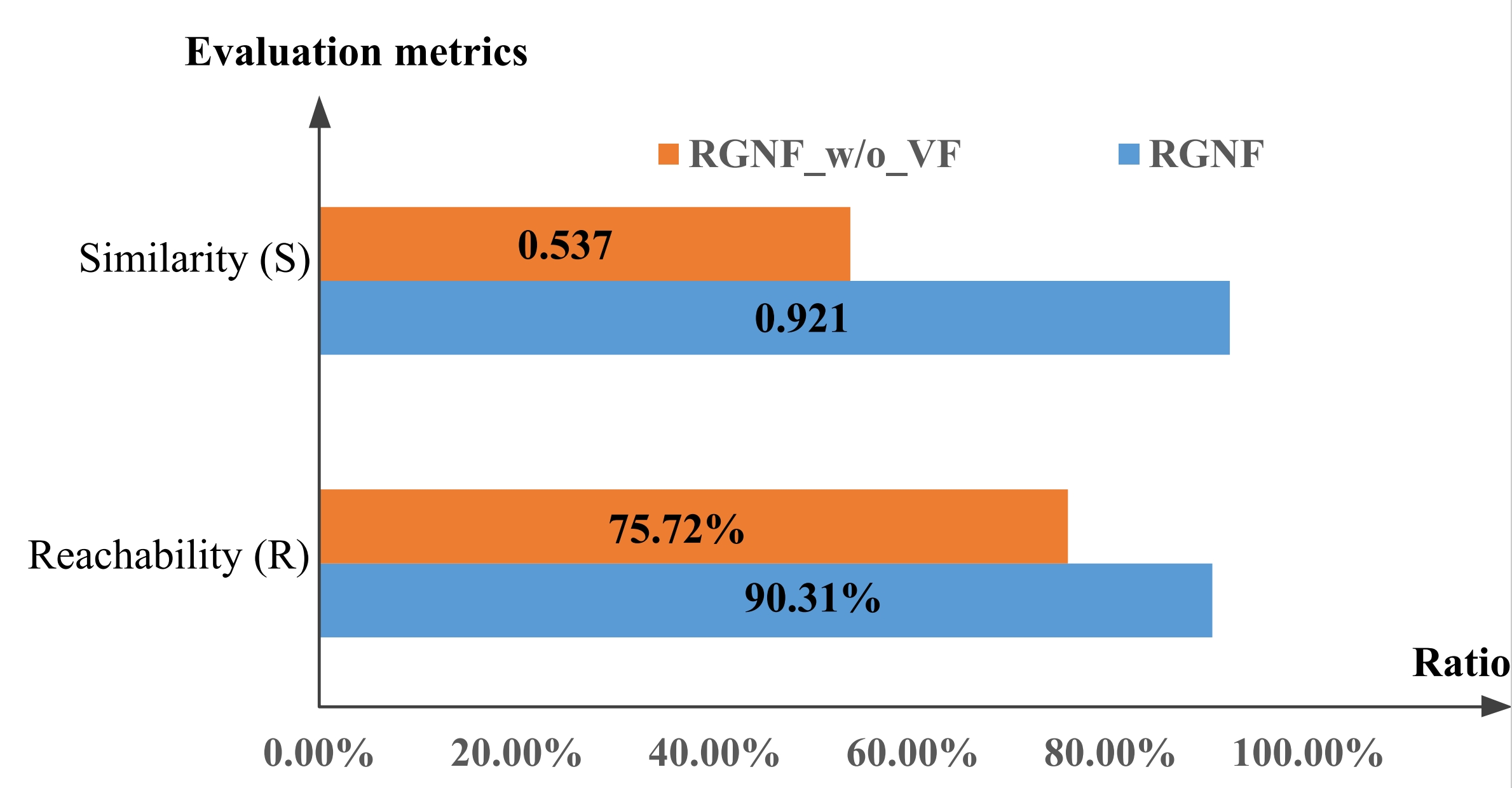}
\caption{The results of ablation study.}
\label{fig: ablation}
\end{figure}
\subsubsection{Discussion on limitation and generalizability}
The limitations of our approach are in the following two aspects. 
First, if none of the components within the GUI can be focused, meaning that the screen reader cannot access any element within the GUI, then our method naturally cannot be applied to such GUIs. 
Second, our method is not applicable to GUIs that only contain text (e.g., e-books). 
The primary reason is that the text information in such applications only needs to be read sequentially, which does not involve the order of components. 
Regarding the generalizability of our method, in addition to the commonly used screen reader for visually impaired users, Talkback, tested in this study, our method can be applied to other screen readers, such as Webex~\footnote{\url{https://www.webex.com/}} and Tatans~\footnote{\url{https://www.tatans.cn/}}. 
This generalizability is due to the fact that our work is not specific to a specific tool, but to the mobile app itself.
By rearranging the components within the GUI and adjusting their navigation order, it would change the focus of the components within the app, and then when any screen reader accesses the app, its navigation order changes. 
However, we also find a failure case in which our method could not be applied to specific in-app reading tools, such as iReader. 
These tools lock the access logic of all content within the apps, making it impossible for our method to modify or interact with the navigation flow.

\subsection{RQ2: Evaluation of Usefulness by User Study}

The purpose of this research question was to discuss whether the GNF redrawn by our method was more in line with the practical feelings of visually impaired users.
Therefore, we conducted a user study catering to visually impaired users.

\subsubsection{Participants recruitment} In the School of Special Education, we initiated the experiment by recruiting visually impaired participants with varying levels of experience in mobile device usage.
Eight participants were selected from a pool of volunteers who had at least 3 years of experience using apps.
Our participants consisted of visually impaired users ranging in age from 20 to 35 years old, encompassing 4 females and 4 males, and they typically relied on assistive tools (i.e., Talkback) to use apps.
Besides, we examined their levels of vision to ensure that any discrepancy in vision would not impact the experimental outcomes. 
Among the 8 participants, all of them were completely blind users to underscore their inability to influence the experimental results through visual means.
Notably, these users who participated in the experiment were re-recruited and were not the same as the users who made the true set in RQ1 and conducted the investigation in Section~\ref{sub: Pre}.
This ensured that participants' assessments of the utility of our method were not influenced by subjective factors stemming from their memorize.

\subsubsection{Preparation} To ensure that the experiment is successfully conducted and to protect the rights of participants, we provided a short training session to these users before the experiment begins.
This training included a general introduction to the experiment and clear instructions on the tasks we expected them to complete.
Participants were encouraged to ask questions and sought clarification on any aspects of the experiment during this training phase, and we made much effort to accommodate their needs and concerns.
Regarding the apps to be assessed by these participants, we presented them with the 25 popular apps collected in RQ1 and encouraged them to collectively discuss and selected apps they had never used before for the experiment.
This was done to mitigate the potential influence of their prior usage experiences on the evaluation results.
As a result, participants selected 8 apps for this experiment that they had never used before.
Furthermore, we strictly adhered to ethical guidelines throughout the study.
We also provided participants with informed consent forms, outlining the purpose of the experiment, data collection procedures, and their rights. 
Confidentiality and anonymity of participants were maintained, and any personal information was securely handled.

\subsubsection{Research settings} This user study comprised the following steps.
First, we randomly divided the 8 participants into two groups, denoted as $g_1$ ($P_1$, $P_2$, $P_3$, and $P_4$) and $g_2$ ($P_5$, $P_6$, $P_7$, and $P_8$).
The apps required for the experiment are also divided into two groups (each containing 4 apps) and further subdivided within these groups based on whether their GNFs undergo redraw using our method.
Let $a_1$ denote the first group of 4 apps that are not GNF redrawn, and $a_2$ denote the redrawn apps; similarly, let $b_1$ represent the latter group of 4 apps that are not GNF redrawn, and $b_2$ is the redrawn apps.
Among them, participants in group $g_1$ were responsible for evaluating the apps in $a_1$ and $b_2$, while participants in group $g_2$ interacted with the apps in $a_2$ and $b_1$.
Participants were assigned the apps in a completely randomized order, and they evaluated them in parallel.
The purpose of such a grouping method is to effectively mitigate the impact of participant biases due to memory, ensuring a more controlled evaluation of the experimental outcomes.
These participants conducted the app evaluations solely based on their personal user experiences in parallel. 
Furthermore, we assigned a staff member to each participant to address any concern that might arise during their assessment of the usability of GNFs. 
Also, the staff member's role was to assist and clarify, without interfering with the participants' evaluations.

These participants rated the GNFs on a scale of $[0,10]$, with higher scores indicating better performance. 
We continuously recorded the rating results from each participant for subsequent quantitative analysis. 
Additionally, following the completion of all evaluations, we organized a brief discussion session in which visually impaired users could share any issues encountered during the assessment process, and provided insights and suggestions they thought needed improvement in these GNFs.
Notably, to gain a comprehensive understanding of user perspectives, we made no restrictions or specifications in this discussion.
The participants were free to express their opinions, and we selected insightful perspectives for analysis from these opinions.

\begin{table}
\centering
\caption{The ratings of user study.}
\label{tab: study}
\begin{tabular}{ll|cccc|cccc}
\toprule
& App & \textbf{$P_1$} & \textbf{$P_2$} & \textbf{$P_3$} & \textbf{$P_4$} & \textbf{$P_5$} & \textbf{$P_6$} & \textbf{$P_7$} & \textbf{$P_8$} \\
\midrule
$a_1$ & OverDrive & 4 & 3 & 5 & 4 & - & - & - & - \\
& MyChart & 3 & 3 & 4 & 4 & - & - & - & - \\
& Wise & 2 & 4 & 4 & 3 & - & - & - & - \\
& Grab & 4 & 4 & 3 & 3 & - & - & - & - \\
\hline
$a_2$ & OverDrive & - & - & - & - & 8 & 7 & 5 & 7 \\
& MyChart & - & - & - & - & 8 & 8 & 8 & 7  \\
& Wise & - & - & - & - & 9 & 8 & 7 & 8  \\
& Grab & - & - & - & - & 8 & 8 & 7 & 8  \\
\hline
$b_1$ & Andev & - & - & - & - & 4 & 3 & 4 & 4 \\
& ESPN & - & - & - & - & 3 & 4 & 4 & 5 \\
& NBC Sports & - & - & - & - & 4 & 4 & 3 & 3 \\
& NewPipe & - & - & - & - & 3 & 3 & 4 & 3 \\
\hline
$b_2$ & Andev & 7 & 6 & 6 & 7 & - & - & - & - \\
& ESPN & 7 & 5 & 8 & 8 & - & - & - & - \\
& NBC Sports & 7 & 8 & 7 & 7 & - & - & - & -  \\
& NewPipe & 5 & 6 & 6 & 7 & - & - & - & -  \\
\bottomrule
\end{tabular}
\end{table}

\subsubsection{Results and analysis} Table~\ref{tab: study} shows the ratings of these two groups of users after using the redrawn GNF and the default GNF, respectively.
The first column represents the names of the apps evaluated by the users, while columns two through nine denote the ratings provided by each user. 
The results reveal a notable improvement, with the average score of the redrawn GNFs surpassing that of the default counterparts by 3.53.
A closer examination of the data underscores this improvement. 
The apps utilizing our redrawn GNFs achieve the highest score of 9 and the lowest score of 6 among the participants. 
In contrast, when using the default GNFs, the highest rating score is 7, while the lowest score drops to as low as 2. 
The fact that the highest score reached 9 in the redrawn GNFs condition demonstrates the potential for our approach to significantly enhance user satisfaction. 
In contrast, the highest score under the default GNFs is limited to 7, indicating a ceiling effect on user experience with the default navigation.
Moreover, the lowest score of 2 in the default GNF condition highlights the substantial usability challenges faced by visually impaired users when using apps without optimized navigation. 
This also suggests that existing apps may not cater to the needs of these visually impaired users.

During our discussion session, participants provided us with two actionable insights.
One is the importance of maintaining consistency between the default navigation and the redrawn version. 
As one of them said, \emph{``You know, abrupt changes in the way we navigate through apps can be quite disorienting. So, here's a suggestion: how about introducing a transition phase? It could help us adapt gradually to the new sequence.''}
While optimizing the sequence is crucial, abrupt changes can be confusing for users who are already familiar with the default navigation. 
It would be helpful to implement a transition phase where users can gradually adapt to the new sequence, ensuring a smoother and less confusing experience during the transition.
Another valuable insight was the need for customizable navigation options, which emerged from the discussion of all participants.
Different users might have unique preferences and habits when interacting with apps. 
Providing the ability for users to customize their navigation sequences based on their specific needs and preferences would greatly enhance the usability of the redrawn GUIs. 
This could involve offering various predefined navigation profiles or even allowing users to create their custom sequences, tailoring the experience to their requirements.

\subsubsection{Future work}
In this work, we focus on a critical subset of app accessibility for visually impaired users, aiming to redraw the GNFs so that these users can operate more accurately.
We believe that it would be fruitful for future studies to explore more kinds of accessibility issues, together with conducting in-depth research on how to fix them.
Enhancing the effectiveness of the redrawn GNF is also a future study worth exploring.
This includes investigating advanced machine learning techniques to dynamically adapt navigation sequences based on user preferences and context.
Exploring additional optimization strategies for complex UI elements such as tables is also a good way to achieve this goal.
Besides, apps in iOS will undoubtedly have accessibility issues, affecting visually impaired users to access information. 
Therefore, we appeal for the related accessibility research across operating systems could be concerned in the future.
This paper calls for future research along this line.

\section{Threats to validity}\label{sec: threats}

The first risk is the datasets we collected and labeled might be insufficient when constructing the ground truth, while more accurate datasets would show better performance.  
However, this also involves the concern that not all apps could be used by visually impaired users. 
For those apps that are not commonly used, the data information would be not worth referring to, while the datasets we collected are as representative as possible. 
Besides, any human-involved experiments would cause threats, the labels of sequences mainly depend on the personal experience of visually impaired users, so the results are highly subjective and it is hard to evaluate whether they are accurate or not. 
To eliminate such subjective implications, we guaranteed that each app was annotated by at least two visually impaired users.
Also, we had addressed our work involves the recruitment of visually impaired users for three separate tasks: investigation, creating the ground truth, and user study. 
If the same group of users is involved in all three tasks, it could introduce subjective bias into the results. 
To mitigate this risk, we recruited new participants for each task to ensure that visually impaired users involved in different activities were distinct.

Another risk is that our reliance on the Gestalt psychological model is driven by a small-scale survey of visually impaired users, which may introduce bias into the conclusions. 
Still, in this survey, we endeavored to recruit participants from diverse professions and age groups to ensure a certain level of representativeness. 
Also, the Gestalt psychological model, as a highly influential model, contributes to the credibility of our approach to a certain extent.
The last risk is that Gestalt psychology also incorporates other two laws, namely, the \emph{law of closure} and the \emph{law of continuity}, to elucidate that people tend to perceive and complete tasks in a continuous and closed manner.
We did not select to incorporate these two laws in guiding the redraw of the GNFs. 
The rationale behind this decision lies in the fact that the information obtained by visually impaired users when utilizing screen readers such as Talkback to access information within the GUI is inherently presented as a continuous sequence. 
Consequently, the GNF derived from this information is also inherently continuous, aligning with the stipulations of Gestalt psychology. 
As such, the inclusion of these two laws is deemed unnecessary and has no impact on the redraw of the GNFs.
Finally, in our selected algorithms, Gaussian blur might sometimes oversimplify the image, potentially causing the loss of important details in certain cases. 
Similarly, Hausdorff distance, while effective for shape comparison, may not capture all aspects of content similarity, which could be a limitation if visual content is also important in the interface. However, we made these choices based on the primary objective of the study: to prioritize structural organization for redrawing the navigation flow of GUIs for visually impaired users.

\section{Related Work}\label{sec: related work}

In this section, we introduced the research related to our work, mainly involving two aspects: GUI accessibility research and the exploration of GUI navigation.

\subsection{GUI accessibility research}

Currently, research on GUI accessibility primarily revolves around two directions. 
One involves empirical explorations to analyze accessibility issues within GUIs, serving as a wake-up call for developers. 
The other entails various tools developed by both industry and academia to identify and pinpoint these issues.

For the first direction, Vendome et al.~\cite{Vendome2019CanEU} conducted a key study on the accessibility of Android apps. 
They found that developers are often more inclined to focus on the needs of ordinary users and neglect the needs of visually impaired users.
In 2021, Alshayban et al.~\cite{Alshayban2020AccessibilityII} further extended this exploration to nearly one million apps in Google Play.
Their findings showed that more than 80\% of these apps have accessibility problems, especially the lack of alternative text, small size, unobvious color contrast and narrow component spacing, which seriously affect the normal operation of visually impaired users.
These studies lay a foundation for the subsequent exploration of GUI accessibility, and provide valuable guidance and inspiration for developers.

In the second direction, both industry~\cite{AccessibilityScanner, MAC, AAAT} and academia~\cite{Liu2020OwlES, Alotaibi2023ScaleFixAA, Alotaibi2021AutomatedRO, Alshayban2022AccessiTextAD} have released and proposed many tools that could check the accessibility issues in GUIs.
In industry, Google released the Google Accessibility Test Framework~\cite{AccessibilityTF} and the corresponding tool called Accessibility Scanner~\cite{AccessibilityScanner} in 2016.
Then, IBM followed with the Mobile Accessibility Checker (MAC)~\cite{MobileAccessibilityChecker} to check the accessibility issues in GUIs.
Also, there are many tools proposed by the researchers, including the PUMA~\cite{Hao2014PUMAPU}, MATE~\cite{Eler2018AutomatedAT}, and XBot~\cite{Chen2021AccessibleON}.
Hao et al.~\cite{Hao2014PUMAPU} developed a programmable UI-automation analysis tool, which could provide significant assistance for visually impaired users.
A similar detection tool is also proposed by Eler et al.~\cite{Eler2018AutomatedAT}, they designed an automated tool, named MATE, which automatically explores apps while applying different checks for accessibility issues related to visual impairment.
Chen et al.~\cite{Chen2021AccessibleON} proposed an Xbot to check the accessibility issues in the GUIs, and made a better performance of collecting accessibility issues.
Furthermore, in our previous research~\cite{10338828}, we have implemented a tool called AccessFixer that could convert GUIs into GUI-Graphs, and then take advantage of the spectrum fluctuation of Relational-Graph Convolutional Neural Network (R-GCN) model~\cite{Schlichtkrull2018ModelingRD} during relationship prediction to fix accessibility issues.

Although existing research and tools well support and improve GUI accessibility, they do not pay attention to and solve the problems of GNFs.
Therefore, compared with these existing tools, our work pays special attention to the issues in the navigation sequence, and further improves the experience of visually impaired users.

\subsection{Exploration of GUI navigation}

The accessibility issues on navigation flow are also crucial factors affecting the visually impaired users to use the apps~\cite{Milne2018Blocks4AllOA, Hara2013CombiningCA, Bigham2010AccessibilityBD}.
To this end, in 2022, Alotaibi et al. proposed a series of tools to detect issues with web keyboard navigation~\cite{Chiou2023DetectingDK, Chiou2023BAGELAA} and mobile application GUI navigation~\cite{Alotaibi2022AutomatedDO}.
These methods and tools provide effective technical support for identifying the key problems in GUI navigation.
Meanwhile, Koutrika et al.~\cite{Koutrika2015GeneratingRO} proposed a framework that could automatically organize a collection of documents in a tree from general to more specific documents.
It also allows a user to choose a reading sequence over the documents.
Another exploration of reading sequences is the Spatio-Temporal Fusion Module (STFM) proposed by Zhang et al.~\cite{Zhang2019SpatioTemporalFB}.
Such a module could well be applied to the navigation flows to maintain the local spatial information and to reduce the feature dimensions.
Meanwhile, this method of utilizing spatial-temporal structure to deal with the sequence issues also inspires us to conduct a similar method that could adapt to the reading sequences of TalkBack.
Further, Salehnamadi et al. proposed the Latte~\cite{Salehnamadi2021LatteUA} that automatically reuses tests written to evaluate an app’s functional correctness to assess its accessibility.
Focusing on eliminating the tediousness of manual accessibility testing, they also introduced Groundhog~\cite{Salehnamadi2022GroundhogAA} and a record-and-replay technique~\cite{Salehnamadi2023AssistiveTechnologyAM}.
These approaches were designed to test and evaluate accessibility issues within the GUI, utilizing intuitive interactive video information and a variety of developer interaction actions, respectively.
Their proposed methods effectively detect accessibility issues within the GUI layout and navigation, but do not provide an appropriate strategy for redrawing the navigation.

Despite the above studies on sequences providing critical technical and inspirational support for our work, they are designed for the sequences in texts or image texts, and are more focused on exploring how to detect issues within the navigation.
Further, there remains a gap in research concerning the navigation sequences specifically utilized by visually impaired users when interacting with screen readers, and hard to provide a method for reconstructing the navigation.

\section{Conclusion}\label{sec: conclusion}
In conclusion, our work addressed the challenge of enhancing the accessibility of GUIs by redrawing their navigation flow. 
We proposed a Gestalt psychology-inspired method, named RGNF, to redraw the GNF and made it more accessible for visually impaired users.
This approach involved utilizing the laws of proximity and similarity to group GUI components into regions and reorganizing component navigation within the grouped regions.
We conducted a series of experiments, including effectiveness assessments and usefulness evaluations, to evaluate our proposed approach. 
The results showed good performance in both effectiveness and usefulness, with visually impaired users finding that our redrawn GNFs were more user-friendly and efficient in app navigation. 
These findings underscore the importance of redrawing GUI navigation flows to the specific needs of visually impaired users and highlight the potential for further advancements in this field.

\section{Acknowledgment}
The work was funded by ``the Fundamental Research Funds for the Central Universities, JLU, 2022-JCXK-16'', ``the National Natural Science Foundation of China (NSFC) No. 62102160'', and supported by ``Jilin Provincial Natural Science Foundation, 20230101070JC''.

%%
%% The next two lines define the bibliography style to be used, and
%% the bibliography file.
\bibliographystyle{IEEEtran}
\bibliography{accessibility}

%%
%% If your work has an appendix, this is the place to put it.

\end{document}